\documentclass[fleqn,10pt]{wlscirep}
\usepackage[T1]{fontenc}  
\usepackage{multirow}
\usepackage{epsfig,epsf,psfrag} 
\usepackage{graphicx} 
\usepackage{subfigure}
\usepackage{amsmath}
\usepackage{hyperref}
\usepackage{pslatex}
\usepackage{epic,eepic} 
\usepackage{color,pstcol}
\usepackage{pstricks} 
\usepackage{fancyhdr}
\usepackage{sidecap}    
 \title{Tuning the 0$-\pi$ Josephson junction with a magnetic impurity: Role of tunnel contacts, exchange coupling, $e-e$ interactions and high-spin states} 
  \author[1]{Subhajit Pal}
\author[1,*]{Colin Benjamin}

\affil[1]{School of Physical Sciences, National Institute of Science Education \& Research, HBNI, Jatni-752050,\ India}

\affil[*]{colin.nano@gmail.com}


\begin{abstract}
We propose Josephson junction with a high-spin magnetic impurity sandwiched between two superconductors. This system shows a $\pi$ junction behavior as a function of the spin magnetic moment state of the impurity, the interface transparency, exchange coupling and electron-electron interactions in the system. The system is theoretically analyzed for possible reason behind the $\pi$ shift. The crucial role of spin flip scattering is highlighted. Possible applications in quantum computation of our proposed tunable high spin magnetic impurity $\pi$ junction is underscored. 
\end{abstract}
\begin{document}
\flushbottom
\maketitle
\section{Introduction}
A tunable $0-\pi$ Josephson junction has inherent potential applications as a cryogenic memory element which is an important component of a superconducting computer which would be much more energy efficient than supercomputers\cite{herr,holmes,ging} based on current semiconductor technology. Further $\pi$ junctions are in high demand as the basic building blocks of a quantum computer\cite{Feo}. In Ref.~\cite{cleu} a Josephson junction in a carbon nanotube setup sandwiched between two superconductors shows a gate-controlled transition from the $0$ to the $\pi$ state. Further in Ref.~\cite{zhu} a superconductor/quantum-dot/superconductor junction is considered and various mechanisms are explored to see the $\pi$-junction transition. In this work, we show that a high spin magnetic impurity(HSM) sandwiched between two s-wave superconductors can transit from a $0$ to $\pi$ Josephson junction via tuning any one of the system parameters like strength of tunnel contact, the spin $S$ or magnetic moment of HSM or the exchange coupling  J. Our motivation for looking at this set up stems from the fact that most of the $\pi$ junction proposals depend on either ferromagnet or d-wave superconductor\cite{Rya,Van} for their functioning.  Integrating Ferromagnets into current superconductor circuit technology hasn't been easy. Controlling Ferromagnets is an onerous task. Further d-wave superconductors, in effect high $T_{c}$ superconductors also have a poor record of being integrated into current superconductor technology. Thus, in this work we obviate the need for any Ferromagnets or d-wave superconductors by implementing a Josephson $\pi$ junction with a magnetic impurity. This magnetic impurity can be an effective model for a spin flipper or even a high spin molecule in certain limits. 

Our paper is organized as follow- in section 2, we introduce our model, give a theoretical background to our study with Hamiltonian, wavefunctions and boundary conditions to calculate the Josephson current. In section 3, we use the Furusaki-Tsukuda formalism to calculate the total Josephson current. To calculate the individual contribution-(i) the bound state we take the derivative of bound state energy with respect to phase difference and (ii) for the continuum contributions use the formalism developed in Refs.~\cite{Bag} and \cite{Kri}. Following this we plot the Andreev bound states as a function of phase difference for different values of spin and magnetic moment of the HSM in section 4. The next section concerns with the Josephson supercurrent plots. We bring out the fact that the $0-\pi$ junction behavior can be tuned via the spin of the HSM. In section 6, we study the Josephson current in the long junction limit and find that the $\pi$ junction is robust to increase in length of normal metal region. Section 7 deals with the free energy of our system and we especially concentrate  on the parameters necessary to exhibit bistable junction behavior, necessary precursor to Josephson qubits. The effects of interface transparency on the $0-\pi$ junction behavior is brought out in section 8. The exchange interactions between the HSM and the electrons in normal metal can also play a crucial role on the tun-ability of $0-\pi$ Josephson junction, this is explored in section 9. We reveal that the electron-electron interactions in our system has a nontrivial role in the tun-ability of the $\pi$ junction in section 10. Section 11 deals with the effect of the HSM spin states on the Josephson supercurrent. These too in conjunction with electron-electron interactions have a nontrivial role in the tun-ability of our Josephson $\pi$ junction. Finally, the paper ends with a perspective on future endeavors.
\section{Theory}
\subsection{Hamiltonian}
The Hamiltonian\cite{AJP,mauri} used to describe a HSM is given by-
\begin{equation}
H_{HSM}=-J_{0}\vec{s}.\vec{S}  
\end{equation}
 The above model for a magnetic impurity in a Josephson junction matches quite well with solid-state scenarios such as
seen in  1D quantum wires or graphene  with an embedded magnetic impurity or quantum dot\cite{palma}. The electrons in the normal metals interact with HSM via the Hamiltonian with just a exchange term $-J_{0}\vec{s}.\vec{S}$, where $J_{0}$ is the strength of the exchange interaction, $\vec{s}$ is the electronic spin and $\vec{S}$ is the spin of the HSM. $J_{0}(=\frac{\hbar^2k_{F}J}{m^{\star}})$, with $J$ being the relative magnitude of the exchange interaction which ranges from $0-3$ in this work, $m^{\star}$ is the electronic mass and Fermi wavevector $k_{F}$ is obtained from the Fermi energy $E_{F}$ which is the largest energy scale in our system$~1000\Delta$, $\Delta$- the superconducting gap for a widely used s-wave superconductor like lead is $1$ meV. Substituting the value of the Fermi wavevector so obtained in the formula for $J_{0}$ we get $J_{0}=0.778$ eV (if $J=2$). {In a realistic HSM there is a anisotropy term\cite{peng}($-DS_{z}^2$) in the Hamiltonian (Eq.~1). The magnitude of \textquotedblleft anisotropy parameter\textquotedblright{} denoted by $D$ is $56$ $\mu$eV. This value has been found by different spectroscopic techniques like Electron Paramagnetic Resonance EPR, neutron scattering and superconducting quantum interference device SQUID magnetometry\cite{Bur}. Thus exchange interaction $J_{0}$ is almost 14000 times larger than anisotropy parameter $D$. Therefore, we can neglect the term $D$ in Hamiltonian of HSM.} 
Our system consists of two normal metals with a HSM sandwiched between two conventional s-wave singlet superconductors. The superconductors are isotropic, and we consider an effective 1D model as shown in Fig.~1, it depicts a HSM at $x=0$ and two superconductors at $x<-a/2$ and $x>a/2$. There are normal metal regions in $-a/2<x<0$ and $0<x<a/2$.
The model Hamiltonian in Bogoliubov-de Gennes formalism of our system is a $4\times4$ matrix which is given below:
\begin{eqnarray}
  \begin{pmatrix}
    H\hat{I} & i\Delta \hat{\sigma}_{y}\\
    -i\Delta^{*} \hat{\sigma}_{y} & -H\hat{I}
  \end{pmatrix} \Psi(x)& =& E \Psi(x),
\end{eqnarray}
$H= p^2/2m^\star + V[\delta(x+a/2)+\delta(x-a/2)]-J_{0}\delta(x)\vec s.\vec S -E_{F}$, here 
$p^2/2m^\star$ is the kinetic energy of an electron with effective mass $m^\star$, $V$ is the strength of the $\delta$ potential at the interfaces between normal metal and superconductor, $J_{0}$ is the strength of exchange interaction between the electron with spin $\vec{s}$ and a HSM with spin $\vec{S}$. Further, $\Psi$ is a four-component spinor, $\hat{\sigma}$ is Pauli spin matrix and $\hat{I}$ is $2\times2$ unit matrix, $E_{F}$ being the Fermi energy. The superconducting gap parameters $\Delta$ for left and right superconductor, are assumed to have the same magnitude but different phases $\varphi_{L}$ and $\varphi_{R}$ and are given by $\Delta=\Delta_{0}[e^{i\varphi_{L}}\theta(-x-a/2)+e^{i\varphi_{R}}\theta(x-a/2)]$, $\theta(x)$ is the Heaviside step function, $\Delta_{0}$ is temperature dependent gap parameter and it follows
$\Delta_{0}\rightarrow \Delta_{0}\tanh(1.74\sqrt{(T_{c}/T-1)})$, where $T_{c}$ is the superconducting critical temperature\cite{annu}.\\
{The wavefunctions for different regions and boundary conditions at different interfaces of our system are given in Supplementary material section I. By imposing the boundary conditions on the wavefunctions one can get the different scattering amplitudes.}
\begin{figure}[h]
\centering{ \includegraphics[width=0.7\textwidth]{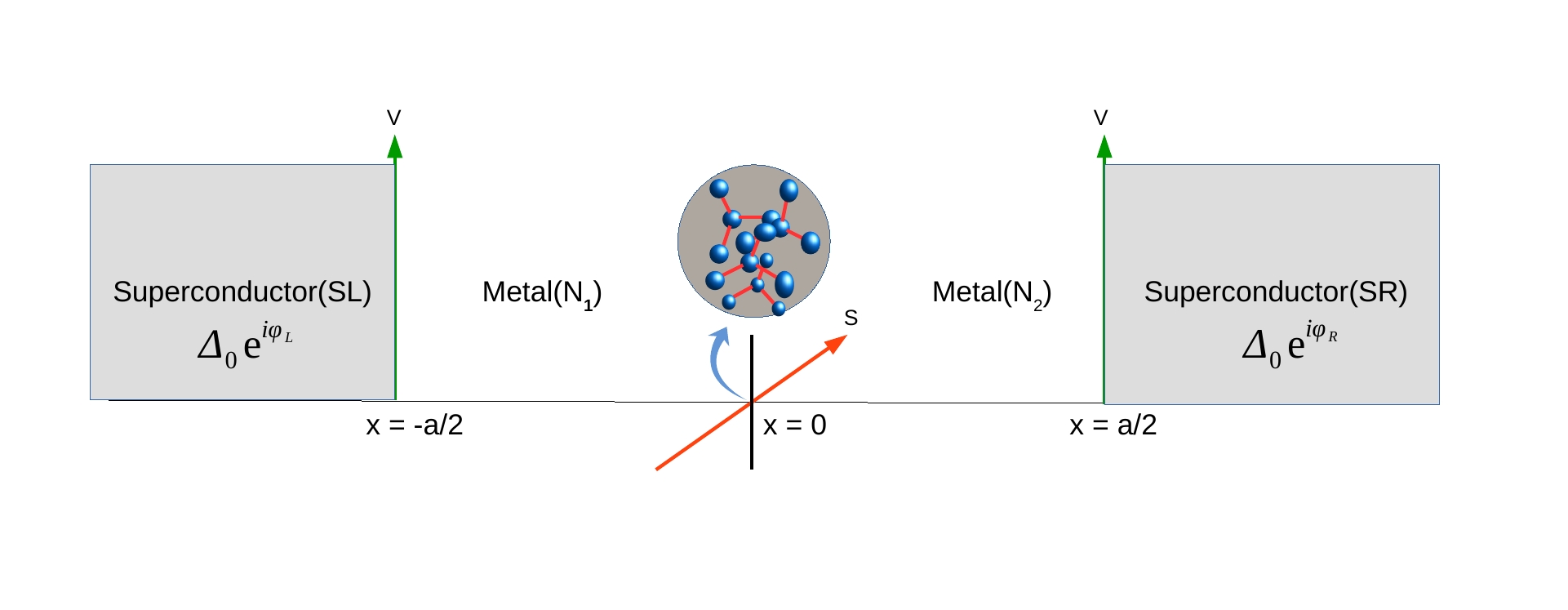}}
\caption{Josephson junction composed of two normal metals and a high spin magnetic impurity with spin $S$ and magnetic moment $m'$ at $x=0$ sandwiched between two s-wave superconductors.}
\end{figure}
\section{Josephson current in presence of a HSM}
\subsection{Total Josephson current}
Using the generalized version of the Furusaki-Tsukuda formalism\cite{Furu} we can calculate the total dc Josephson current-
\begin{equation}
I_{T}(\varphi)=\frac{e\Delta_{0}}{2\beta\hbar}\frac{1}{2\pi}\int_{0}^{2\pi}\sum_{\omega_{n}}\frac{q_{+}(\omega_{n})+q_{-}(\omega_{n})}{\Omega_{n}}\times\Bigg[\frac{a_{1}(\omega_{n})-a_{2}(\omega_{n})}{q_{+}(\omega_{n})}+\frac{a_{3}(\omega_{n})-a_{4}(\omega_{n})}{q_{-}(\omega_{n})}\Bigg]d(k_{F}a),
\end{equation}
herein $\omega_{n}=(2n+1)\pi/\beta$ are fermionic Matsubara frequencies with $n=0,\pm1,\pm2,...$ and $\Omega_{n}=\sqrt{\omega_{n}^2+\Delta_{0}^2}$. $k_{F}a$ is the phase accumulated in normal metal region. $\beta=1/kT$ is the inverse temperature. $q_{+}(\omega_{n}),q_{-}(\omega_{n})$, and $a_{i}(\omega_{n})$ are obtained from $q_{+},q_{-}$ and $a_{i}$ by analytically continuing $E$ to $i\omega_{n}$. Here $a_{i}(i=1,2,3,4)$ with $a_{1}=r_{eh}^{\uparrow\downarrow}$ is the Andreev reflection coefficient without flip for electron up incident in left superconductor, similarly $a_{2}=r_{eh}^{\downarrow\uparrow}$ is the Andreev reflection coefficient without flip for electron down incident in left superconductor, $a_{3}=r_{he}^{\uparrow\downarrow}$ and $a_{4}=r_{he}^{\downarrow\uparrow}$ are the Andreev reflection coefficients without flip for hole up and hole down incident in left superconductor respectively. There are other ways of writing the Josephson supercurrent formula in Furusaki-Tsukuda approach\cite{Costa,LINDER}, all such ways give identical total Josephson current. These different ways involve different scattering amplitudes, as due to the fact that Furusaki-Tsukuda procedure obeys both detailed balance as well as probability conservation, allowing for the possibility of different representations of the same formula. We sum over the Matsubara frequencies numerically. The detailed balance conditions\cite{Furu} are verified as follows:
\begin{equation}
\frac{a_{1}(-\varphi,E)}{q_{+}}=\frac{a_{4}(\varphi,E)}{q_{-}}, \frac{a_{2}(-\varphi,E)}{q_{+}}=\frac{a_{3}(\varphi,E)}{q_{-}}\nonumber 
\end{equation}
\subsection{Bound state contribution}
Neglecting the contribution from incoming quasiparticle\cite{LINDER} and inserting the wave function into the boundary conditions, we get a homogeneous system of 24 linear equations for the scattering amplitudes. If we express the scattering amplitudes in the two normal metal regions by the scattering amplitudes in the left and right superconductor we get a homogeneous system of 8 linear equations\cite{annu},
\begin{equation}
 Mx=0
\end{equation}
where $x$ is a $8\times1$ column matrix and is given by $x=[r_{ee}^{\uparrow\uparrow},r_{ee}^{\uparrow\downarrow},r_{eh}^{\uparrow\uparrow},r_{eh}^{\uparrow\downarrow},t_{ee}^{\uparrow\uparrow},t_{ee}^{\uparrow\downarrow},t_{eh}^{\uparrow\uparrow},t_{eh}^{\uparrow\downarrow}]$ and $M$ is a $8\times8$ matrix which is explicitly written in {supplementary material section II.} For a nontrivial solution of this system of equations, det $M=0$, we can get a relation between the Andreev bound state energy and phase difference, i.e., Andreev levels with dispersion $E_{i}$, $i= \{1,...,4\}$\cite{Been}. We find that $E_{i}(\varphi)=E_{\sigma}^{\pm}(\varphi)=\pm E_{\sigma}(\varphi), (\sigma=\uparrow,\downarrow)$ and 
\begin{equation}
E_{\sigma}^{\pm}(\varphi)=\pm\Delta_{0}\sqrt{\frac{|A(\varphi)|+\rho_{\sigma}\sqrt{|B(\varphi)|}}{2|C|}}\label{Eq:bb} 
\end{equation}
wherein $\rho_{\uparrow(\downarrow)}=+1(-1)$ and $A(\varphi),B(\varphi),C$ depend on all junction parameters. Their explicit form is given in {supplementary material section III.} For simplicity we have taken all wavevectors equal to the Fermi wavevector (Andreev approximation). For transparent regime ($Z=0$) we find-
\begin{align}
\begin{split} 
E_{\sigma}^{\pm}(\varphi)={} & \pm\frac{\Delta_{0}}{\sqrt2}                                                                                                         (\surd((2(8+J^4(F_{2}^2+m'+m'^2)^2+J^2(3+2F_{2}^2+6m'(1+m'))+(8+J^2(1-2F_{2}^2+2m'(1+m')))\cos(\varphi))\\                                                                                                    
                             &+\rho_{\sigma}\surd(2J^2(64F_{2}^4J^2+3(J+2Jm')^2+4F_{2}^2(16+J^2(5+4m'(1+m')))+4J^2(-4F_{2}^2+16F_{2}^4-(1+2m')^2)\cos(\varphi)\\
                             &+((J+2Jm')^2-4F_{2}^2(16+(J+2Jm')^2))\cos(2\varphi))))/(16+
                             J^4(F_{2}^2+m'+m'^2)^2+J^2(4+8F_{2}^2+8m'(1+m')))))\label{Eq:eb}
\end{split}
\end{align}
For $Z=0$, interestingly the bound states are independent of any phase ($k_{F}a$) accumulated in normal metal region.
For tunneling regime ($Z\rightarrow Large$) we get-
\begin{equation}
E_{\sigma}^{\pm}(\varphi)=\pm\Delta_{0}\Bigg[1+\frac{(8+J^2(1-2F_{2}^2+2m'(1+m'))+(8+J^2(-1+2F_{2}^2-2m'(1+m')))\cos(k_{F}a))\cos(\varphi)}{16Z^4\sin(\frac{k_{F}a}{2})^2(4 -J^2(F_{2}^2+m'+m'^2)+(4+J^2(F_{2}^2+m'+m'^2))\cos(k_{F}a)+2J\sin(k_{F}a))^2}\Bigg]
\end{equation}
For $Z\rightarrow Large$, we can clearly say that bound states are phase ($k_{F}a$) dependent. 
From Andreev bound states energies Eq. \ref{Eq:bb} we can derive the Josephson bound state current\cite{Golu}- 
\begin{equation}
I_{B}(\varphi) = \frac{2e}{\hbar}\frac{1}{2\pi}\int_{0}^{2\pi}\sum_{i}f(E_{i})\frac{dE_{i}}{d\varphi}d(k_{F}a)
               = -\frac{2e}{\hbar}\frac{1}{2\pi}\int_{0}^{2\pi}\sum_{\sigma}\tanh\Big(\frac{\beta E_{\sigma}}{2}\Big)\frac{dE_{\sigma}}{d\varphi}d(k_{F}a)
\end{equation}               
wherein $e$ is the electronic charge and $f(E_{i})$ denotes the Fermi-Dirac distribution function.
For transparent regime ($Z=0$) we obtain the current-phase relation
\begin{equation} 
\frac{I_{B}(\varphi)}{I_{0}}=\frac{\Delta_{0}\sin(\varphi)((C_{1}+C_{2})\frac{1}{E_{\downarrow}}\tanh\Big(\frac{\beta E_{\downarrow}}{2}\Big)-(C_{1}-C_{2})\frac{1}{E_{\uparrow}}\tanh\Big(\frac{\beta E_{\uparrow}}{2}\Big))}{C_{3}}
\end{equation}
where
\begin{align}
\begin{split}
C_{1}={}& ((\sqrt{2}J^4(4F_{2}^2-16F_{2}^4+(1+2m')^2)+\sqrt{2}J^2(4F_{2}^2(16+(J+2Jm')^2)-(J+2Jm')^2)\cos(\varphi))\\\nonumber
\end{split}\\\nonumber
\begin{split}
C_{2}={}& (8+J^2-2J^2F_{2}^2+2J^2m'(1+m'))\surd(J^2(64F_{2}^4J^2+3(J+2Jm')^2+4F_{2}^2(16+J^2(5+4m'(1+m')))\\
        & +4J^2(-4F_{2}^2+16F_{2}^4-(1+2m')^2)\cos(\varphi)+((J+2Jm')^2-4F_{2}^2(16+(J+2Jm')^2))\cos(2\varphi)))\\\nonumber 
\end{split}\\
\begin{split}
C_{3}={}& ((16+J^4(F_{2}^2+m'+m'^2)^2+J^2(4+8F_{2}^2+8m'(1+m')))\surd(J^2(64F_{2}^4J^2+3(J+2Jm')^2+4F_{2}^2(16\\\nonumber
        & +J^2(5+4m'(1 +m')))+4J^2(-4F_{2}^2+16F_{2}^4-(1+2m')^2)\cos(\varphi)+((J+2Jm')^2-4F_{2}^2(16+(J+2Jm')^2))\cos(2\varphi))))\\\nonumber 
\end{split}
\end{align}

$I_{0}=e\Delta_{0}/\hbar$ and $E_{\uparrow(\downarrow)}$ is given in Eq.~\ref{Eq:eb}. For tunneling regime ($Z\rightarrow Large$) and at $T=0$ we find 
\begin{equation}
\frac{I_{B}(\varphi)}{I_{0}}=\frac{1}{2\pi}\int_{0}^{2\pi}\Bigg[\frac{(8+J^2(1-2F_{2}^2+2m'(1+m'))+(8+J^2(-1+2F_{2}^2-2m'(1+m')))\cos(k_{F}a)) \sin(\varphi)}{4Z^4\sin(\frac{k_{F}a}{2})^{2}(4-J^2(F_{2}^2+m'+m'^2)+(4+J^2(F_{2}^2+m'+m'^2))\cos(k_{F}a)+2J\sin(k_{F}a))^2}\Bigg]d(k_{F}a) 
\end{equation}
\subsection{Continuum contribution}
The continuum contribution to the Josephson current is the collection of currents carried by both electron-like and hole-like quasiparticles outside the gap. Using the formalisms developed earlier in Refs.~[\cite{Bag}, \cite{Tsu}] the continuum contribution from electron-like excitations is given below.\cite{Bag} 
\begin{equation}
I_{C}^{e}(\varphi)=\frac{2e}{h}\frac{1}{2\pi}\int_{0}^{2\pi}(\int_{-\infty}^{-\Delta_{0}}+\int_{\infty}^{\Delta_{0}})\frac{1}{\mid u^2-v^2 \mid }\times[T_{L\rightarrow R}^{e \uparrow\uparrow}(E,\varphi)+T_{L\rightarrow R}^{e \uparrow\downarrow}(E,\varphi)-T_{L\leftarrow R}^{e \uparrow\uparrow}(E,\varphi)-T_{L\leftarrow R}^{e \uparrow\downarrow}(E,\varphi)]f(E)dE d(k_{F}a)\label{Eq:ICe}
\end{equation}
Similarly the continuum contribution from hole-like excitations can be calculated by replacing $'e'$ in Eq. \ref{Eq:ICe} by $'h'$.
In Eq. \ref{Eq:ICe} $T_{L\rightarrow R}^{e \uparrow\uparrow}=|t_{ee}^{\uparrow\uparrow}|^{2}-|t_{eh}^{\uparrow\uparrow}|^{2}$ is the transmission without flip for the electronic currents moving from left to right of the system as depicted in Fig.~1. $T_{L\rightarrow R}^{e \uparrow\downarrow}=|t_{ee}^{\uparrow\downarrow}|^{2}-|t_{eh}^{\uparrow\downarrow}|^{2}$ is the transmission with flip for the electronic currents moving from left to right of the system and similarly $T_{L\leftarrow R}^{e \uparrow\uparrow}$ and $T_{L\leftarrow R}^{e \uparrow\downarrow}$ are the transmission without flip and with flip for the electronic currents moving from right to left of the system respectively. Here we have
\begin{equation}
T_{L\leftarrow R}^{e \uparrow\uparrow}(E,\varphi)=T_{L\rightarrow R}^{e \uparrow\uparrow}(E,-\varphi), T_{L\leftarrow R}^{e \uparrow\downarrow}(E,\varphi)=T_{L\rightarrow R}^{e \uparrow\downarrow}(E,-\varphi)\nonumber
\end{equation}
The hole continuum contribution is found to be equal to the electronic continuum contribution. Therefore, the total continuum current due to electron-like and hole-like excitations is given as follows:
\begin{equation}
 I_{C}(\varphi)=\frac{I_{C}^{e}(\varphi)+I_{C}^{h}(\varphi)}{2}=I_{C}^{e}(\varphi)
\end{equation}
In our work we have verified the total current conservation-$I_{T}(\varphi)=I_{B}(\varphi)+I_{C}(\varphi)$
\begin{figure}[h]  
\includegraphics[width=.9\textwidth]{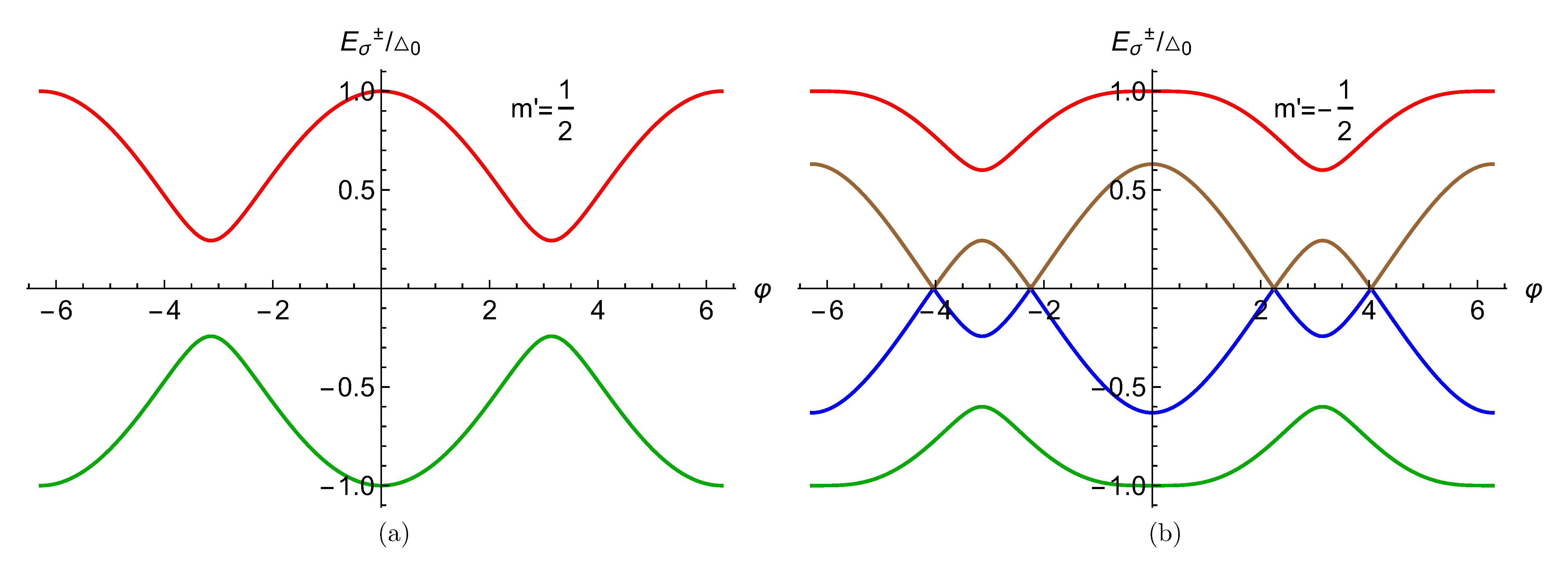}
\caption{\small \sl Andreev bound states as a function of phase difference ($\varphi$). Parameters are $\Delta_{0}=1 meV$, $S=1/2$, $m'=\pm1/2$, $J=1$,  $Z=0$.}
\end{figure}
\section{Andreev bound states}
The Andreev bound states (ABS) as obtained in Eq. \ref{Eq:bb} are analyzed in this section. We focus on the role of spin $S$ and magnetic moment $m'$ of the HSM on ABS. In Fig.~2(a), we plot ABS for $S=1/2$ and $m'=1/2$, as here the spin flip probability $F_{2}=\sqrt{(S-m')(S+m'+1)}=0$ which corresponds to no flip, we get only two bound states, but in Fig.~2(b) with $S=1/2$ and $m'=-1/2$, $F_{2}\neq 0$ thus due to spin flip processes we get four bound states. To address the situation of large spin $S$ in HSM

in Fig.~3 we plot ABS for $S=9/2$ and all allowed $m'$ values. For particular $S$, as $m'$ changes, separation between electron (positive) bound states and hole (negative) bound states increases. Similarly for particular $m'$ as we change $S$, this separation increases. For large $S$, ABS lie at the gap edge. This is seen for large $m'$ as well. This behavior is also seen as one changes $J$, $Z$ as well. We only plot ABS for $m'=\pm1/2,\pm3/2,\pm 9/2$, but we do not plot for $m'=\pm 5/2,\pm 7/2$ because the separation between electron bound states and hole bound states increases from $m'=3/2$ to $m'=9/2$ and these $m'$ values lie between $m'=3/2$ and $m'=9/2$. Large $S$, $m'$, $Z$, $J$ lead to ABS shifting to gap edge.
\begin{figure}[h]  
\includegraphics[width=0.9\textwidth]{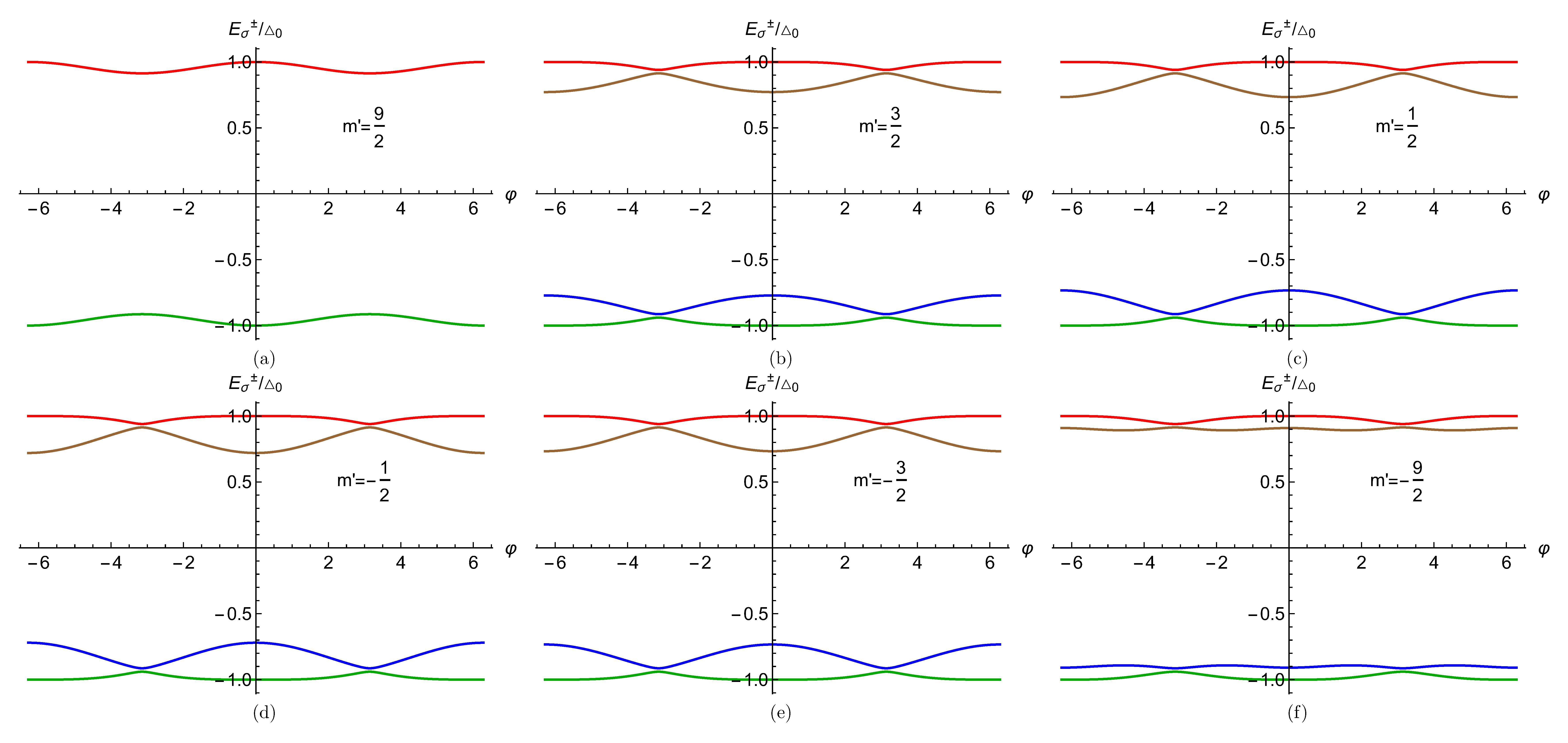}
\caption{\small \sl Andreev bound states as a function of phase difference ($\varphi$). Parameters are $\Delta_{0}=1 meV$, $S=9/2$, $m'=\pm1/2,\pm3/2,\pm9/2$, $J=1$, $Z=0$.}
\end{figure}
\begin{figure}[h]  
\includegraphics[width=.93\textwidth]{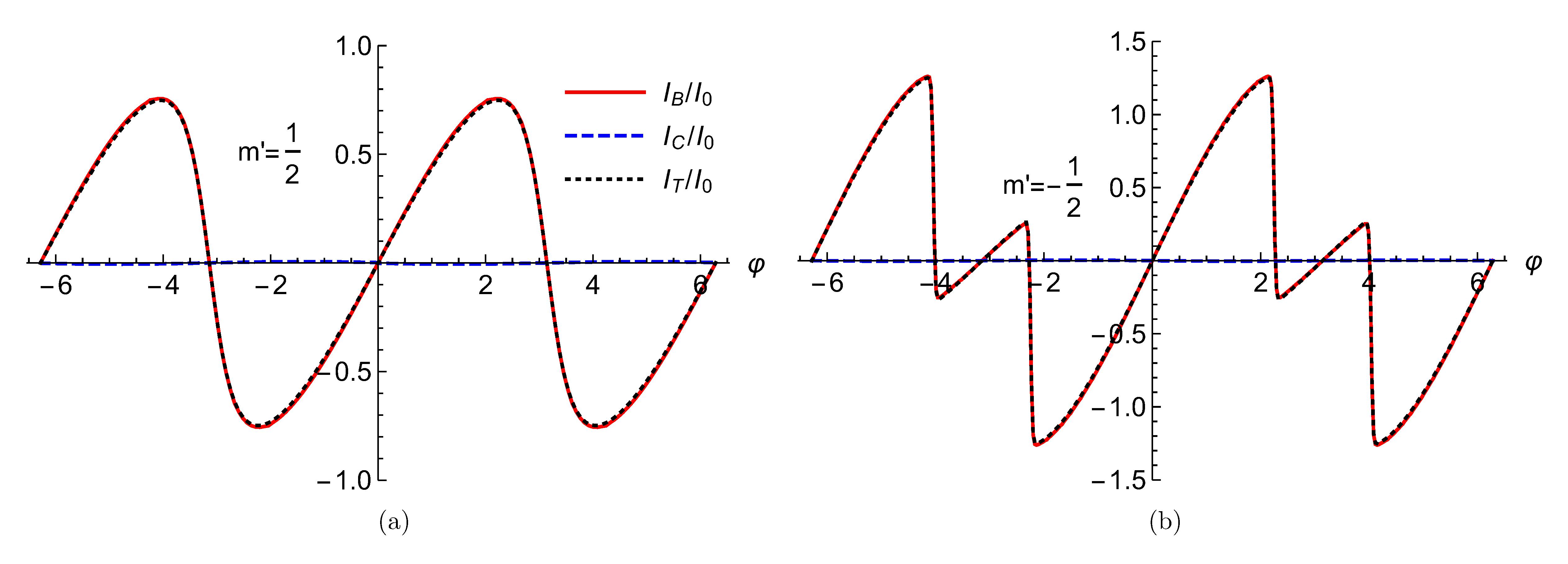}
\caption{\small \sl The bound, continuum and total Josephson current as a function of phase difference ($\varphi$). Parameters are $\Delta_{0}=1 meV$,  $T/T_{c}=0.01$, $S=1/2$, $m'=\pm1/2$, $J=1$, $Z=0$.}
\label{IB}
\end{figure}
\section{Josephson current: $\pi$ junction}
\begin{figure}[h]  
\includegraphics[width=.8\textwidth]{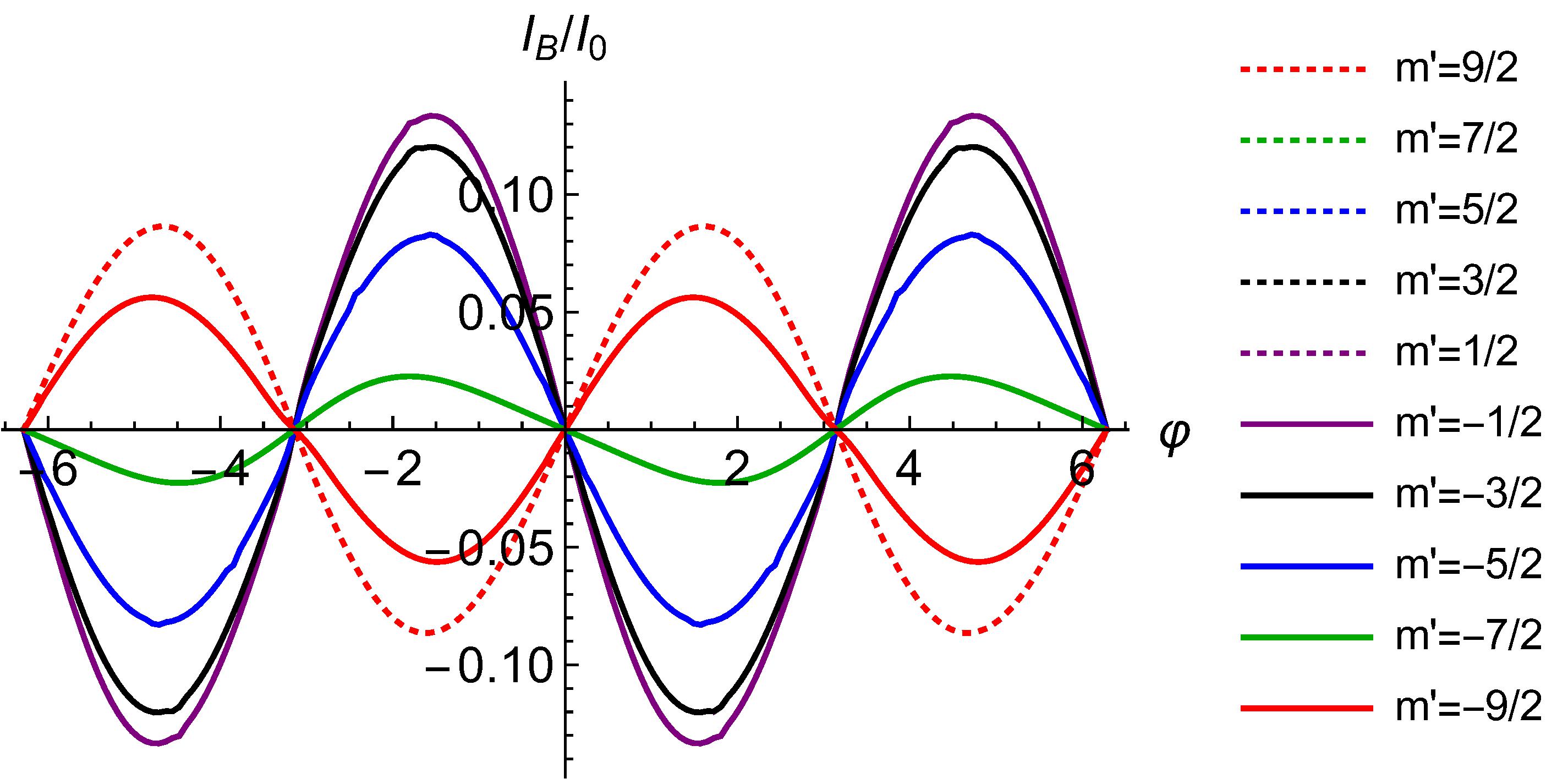}
\caption{\small \sl Josephson supercurrent as a function of phase difference ($\varphi$). Parameters are $\Delta_{0}=1 meV$, $T/T_{c}=0.01$, $S=9/2$, $m'=\pm9/2,\pm7/2,\pm5/2,\pm3/2,\pm1/2$, $J=1$, $Z=0$. Josephson supercurrent for $m'=7/2$ and $m'=-9/2$ are same and similarly for $m'=5/2$ and $m'=-7/2$, $m'=3/2$ and $m'=-5/2$, $m'=1/2$ and $m'=-3/2$ are same.}
\label{IB2}
\end{figure}

The considered model shows $\pi$ junction behavior. To see this, we plot the bound state, continuum and total Josephson currents for $S=1/2$ (Fig.~\ref{IB}) and $S=9/2$ (Fig.~\ref{IB2}). We choose the transparent regime ($Z=0$) case. A separate section will be denoted to effect of tunnel contacts. One can clearly conclude that the continuum contribution of the total current is very small, therefore the bound current and total current are almost same. In Fig.~\ref{IB}(a) as there is no flip we have $0$ junction. For spin flip case, the Josephson current changes sign in $0<\varphi<\pi$ regime.
In Fig.~\ref{IB2} we concentrate on high spin ($S=9/2$) of HSM. Here we also see that for $m'=5/2,3/2,1/2,-1/2,-3/2,-5/2,-7/2$ we get $\pi$ junction. But for $m'=9/2,7/2,-9/2$ we get $0$ junction. So here also there will be a switching from $0$ to $\pi$ and again from $\pi$ to $0$     with change of $m'$ from $9/2$ to $-9/2$. Thus, one can conclude that all $\pi$ shifts are due to spin flip scattering ($F_{2}\neq0$), however the reverse is not necessarily true. This $\pi$-junction state has been studied earlier in Ref.~\cite{Kalen} with spin-active normal metal superconductor (NS) interfaces, but they did not consider any high spin magnetic impurity. Their system shows a $0-\pi$ transition as a function of the kinematic phase, misorientation angle and temperature.   
\section{Josephson Current: Long junction limit}
\begin{figure}[h]  
\includegraphics[width=0.99\textwidth]{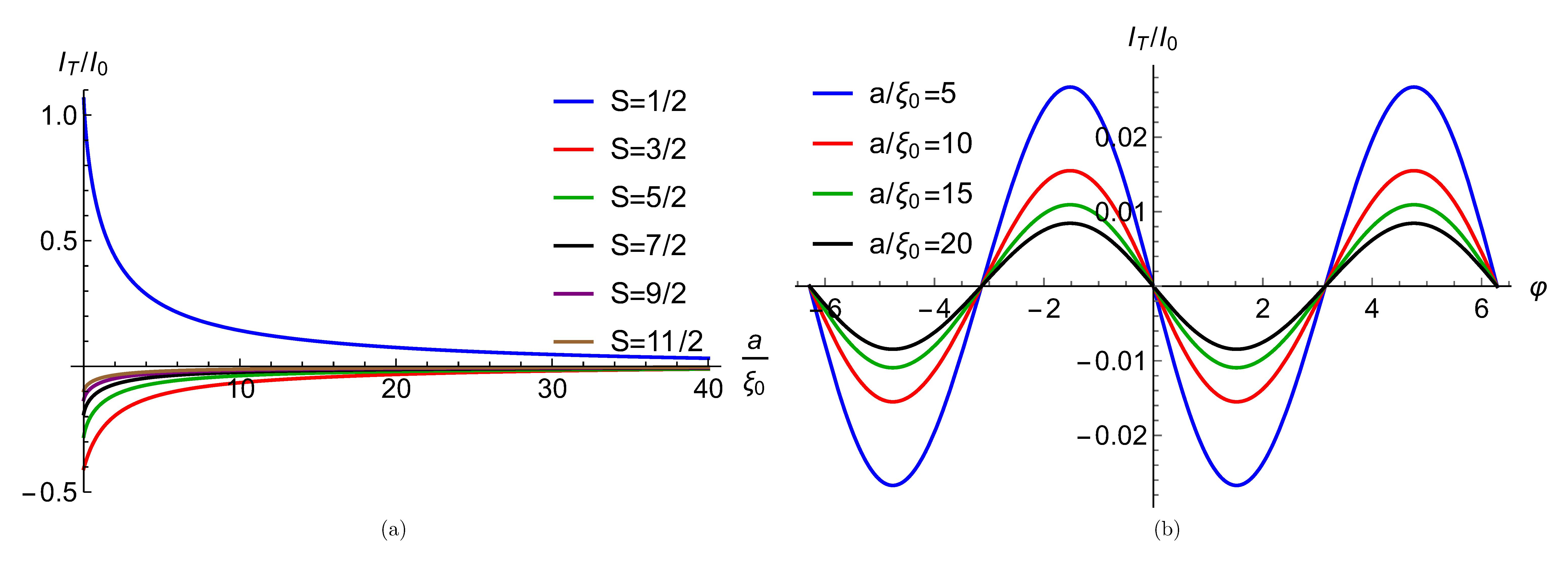}
\caption{\small \sl (a) Josephson supercurrent as a function of junction length ($a$) for different values of spin ($S$) of HSM. Parameters are  $\Delta_{0}=1 meV$,  $T/T_{c}=0.01$, $\varphi=\pi/2$, $m'=-1/2$, $J=1$, $Z=0$, (b) Josephson supercurrent as a function of phase difference ($\varphi$) for different junction length ($a$). Parameters are $\Delta_{0}=1 meV$, $T/T_{c}=0.01$, $S=9/2$, $m'=-1/2$,  $J=1$, $Z=0$ }
\label{IJL}
\end{figure}
There are eight different types of quasiparticle injection into our system: an electron-like quasiparticle (ELQ) with spin up or down or a hole-like quasiparticle (HLQ) with spin up or down injected from either the left or from the right superconducting electrode. 
Following the procedure established in {supplementary material section I.A}, we write the wavefunction for the injection of spin up electron in left side superconductor as-
\begin{eqnarray}
\psi_{SL}(x)=\begin{pmatrix}u\\
                              0\\
                              0\\ 
                              v
                              \end{pmatrix}e^{iq_{+}x}\phi_{m'}^{S}+r_{ee}^{\uparrow\uparrow}\begin{pmatrix}
                              u\\
                              0\\
                              0\\
                              v
                             \end{pmatrix}e^{-iq_{+}x}\phi_{m'}^{S}+r_{ee}^{\uparrow\downarrow}\begin{pmatrix}
                             0\\
                             u\\
                            -v\\
                             0
                             \end{pmatrix}e^{-iq_{+}x}\phi_{m'+1}^{S}+r_{eh}^{\uparrow\uparrow}\begin{pmatrix}
                             0\\
                            -v\\
                             u\\
                             0
                             \end{pmatrix}e^{iq_{-}x}\phi_{m'+1}^{S}+r_{eh}^{\uparrow\downarrow}\begin{pmatrix}
                             v\\
                             0\\
                             0\\
                             u
                             \end{pmatrix}e^{iq_{-}x}\phi_{m'}^{S},\nonumber\\\mbox{for $x<-\frac{a}{2}$}\nonumber
                             \end{eqnarray}
Similarly the corresponding wave function for the right side superconductor is-
\[\psi_{SR}(x)=t_{ee}^{\uparrow\uparrow}\begin{pmatrix}
                              ue^{i\varphi}\\
                              0\\
                              0\\
                              v
                             \end{pmatrix}e^{iq_{+}x}\phi_{m'}^{S}+t_{ee}^{\uparrow\downarrow}\begin{pmatrix}
                             0\\
                             ue^{i\varphi}\\
                             -v\\
                             0
                             \end{pmatrix}e^{iq_{+}x}\phi_{m'+1}^{S}+t_{eh}^{\uparrow\uparrow}\begin{pmatrix}
                             0\\
                             -ve^{i\varphi}\\
                             u\\
                             0
                             \end{pmatrix}e^{-iq_{-}x}\phi_{m'+1}^{S}+t_{eh}^{\uparrow\downarrow}\begin{pmatrix}
                             ve^{i\varphi}\\
                             0\\
                             0\\
                             u
                             \end{pmatrix}e^{-iq_{-}x}\phi_{m'}^{S},\mbox{for $x>\frac{a}{2}$}\]
The wavefunction in the normal metal region ($N_{1}$) is given by for the long junction limit following Ref.~\cite{book},
\begin{eqnarray}
\psi_{N_{1}}(x)=ee^{ik_{e}(x+a/2)}\begin{pmatrix}
                                  v\\
                                  0\\
                                  0\\
                                  0
                                  \end{pmatrix}\phi_{m'}^{S}+fe^{-ik_{e}x}\begin{pmatrix}
                                                                           u\\
                                                                           0\\
                                                                           0\\
                                                                           0
                                                                          \end{pmatrix}\phi_{m'}^{S}+e^{\prime}e^{ik_{e}(x+a/2)}\begin{pmatrix}
                                                                                                                                 0\\
                                                                                                                                 -v\\
                                                                                                                                 0\\
                                                                                                                                 0
                                                     \end{pmatrix}\phi_{m'+1}^{S}+f^{\prime}e^{-ik_{e}x}\begin{pmatrix}
                                                                                                        0\\
                                                                                                        u\\
                                                                                                        0\\
                                                                                                        0
                                                                                                        \end{pmatrix}\phi_{m'+1}^{S}
\nonumber\\+ge^{-ik_{h}(x+a/2)}\begin{pmatrix}
                               0\\
                               0\\
                               -v\\
                               0
                               \end{pmatrix}\phi_{m'+1}^{S}+he^{ik_{h}x}\begin{pmatrix}
                                                                         0\\
                                                                         0\\
                                                                         u\\
                                                                         0
                                                                         \end{pmatrix}\phi_{m'+1}^{S}
                                                                         +g^{\prime}e^{-ik_{h}(x+a/2)}\begin{pmatrix}
                                                                                                       0\\
                                                                                                       0\\
                                                                                                       0\\
                                                                                                       v
                                                                                                      \end{pmatrix}\phi_{m'}^{S}
+h^{\prime}e^{ik_{h}x}\begin{pmatrix}
                      0\\
                      0\\
                      0\\
                      u
                      \end{pmatrix}\phi_{m'}^{S},\mbox{for $-\frac{a}{2}<x<0$}\nonumber
                      \end{eqnarray}
Similarly the wavefunction in the normal metal region ($N_{2}$) is given by-
\begin{eqnarray}
\psi_{N_{2}}(x)=a_{0}e^{ik_{e}x}\begin{pmatrix}
                                 v\\
                                 0\\
                                 0\\
                                 0
                                \end{pmatrix}\phi_{m'}^{S}+be^{-ik_{e}(x-a/2)}\begin{pmatrix}
                                                                              u\\
                                                                              0\\
                                                                              0\\
                                                                              0
                                                                              \end{pmatrix}\phi_{m'}^{S}+a^{\prime}e^{ik_{e}x}\begin{pmatrix}
                                                                                                                               0\\
                                                                                                                               -v\\
                                                                                                                               0\\
                                                                                                                               0
                                                                                                                              \end{pmatrix}\phi_{m'+1}^{S}+b^{\prime}e^{-ik_{e}(x-a/2)}\begin{pmatrix}
                                       0\\
                                       u\\
                                       0\\
                                       0
                                      \end{pmatrix}\phi_{m'+1}^{S}\nonumber\\+ce^{-ik_{h}x}\begin{pmatrix}
                                                                                            0\\
                                                                                            0\\
                                                                                            -v\\
                                                                                            0
                                                                                           \end{pmatrix}\phi_{m'+1}^{S}
+de^{ik_{h}(x-a/2)}\begin{pmatrix}
                   0\\
                   0\\
                   u\\
                   0
                   \end{pmatrix}\phi_{m'+1}^{S}+c^{\prime}e^{-ik_{h}x}\begin{pmatrix}
                                                                      0\\
                                                                      0\\
                                                                      0\\
                                                                      v
                                                                      \end{pmatrix}\phi_{m'}^{S}
                                                                      +d^{\prime}e^{ik_{h}(x-a/2)}\begin{pmatrix}
                                                                                                   0\\
                                                                                                   0\\
                                                                                                   0\\
                                                                                                   u
                                                                                                   \end{pmatrix}\phi_{m'}^{S},\mbox{for $0<x<\frac{a}{2}$}\nonumber
                                                    \end{eqnarray}
For $\rvert E \rvert<<E_{F}$, we can write $k_{e,h}\approx k_{F}\pm \frac{E}{2\Delta_{0}\xi_{0}}$, where $\xi_{0}=E_{F}/(k_{F}\Delta_{0})$ is the BCS coherence length\cite{Kri}. By using the boundary conditions mentioned in {supplementary material section I.B} one can get the different scattering amplitudes. The wavefunction for the other seven types of quasiparticle injection process are constructed in the same way. Using the generalized version of Furusaki-Tsukuda Josephson current formula mentioned in section 3.1 we can calculate the total dc Josephson current for long junction limit. In Fig.~\ref{IJL} we plot the Josephson current for a long junction. In Fig.~\ref{IJL}(a) we plot Josephson supercurrent as a function of junction length $a$ for $\varphi=\pi/2$ and different values of spin ($S$) of HSM from $S=1/2$ to $S=11/2$. We see that Josephson supercurrent dies monotonically with increase of length ($a$) of the junction. For large $a$ the Josephson supercurrent goes to zero. In Fig.~\ref{IJL}(b) we have plotted Josephson supercurrent as a function of phase difference ($\varphi$) for different junction length $a$ and high spin of HSM ($S=9/2$). We see that Josephson supercurrent decreases with increase of junction length $a$. In Fig.~\ref{IJL}(a) and \ref{IJL}(b) the magnetic moment of HSM $m'=-1/2$ and the junction transparency $Z=0$. However, change in length has no effect on the sign of Josephson current. Thus signifying that the $\pi$ junction is robust to change in normal metal length.
\section{Free energy}
\begin{figure}[h]  
\includegraphics[width=.65\textwidth]{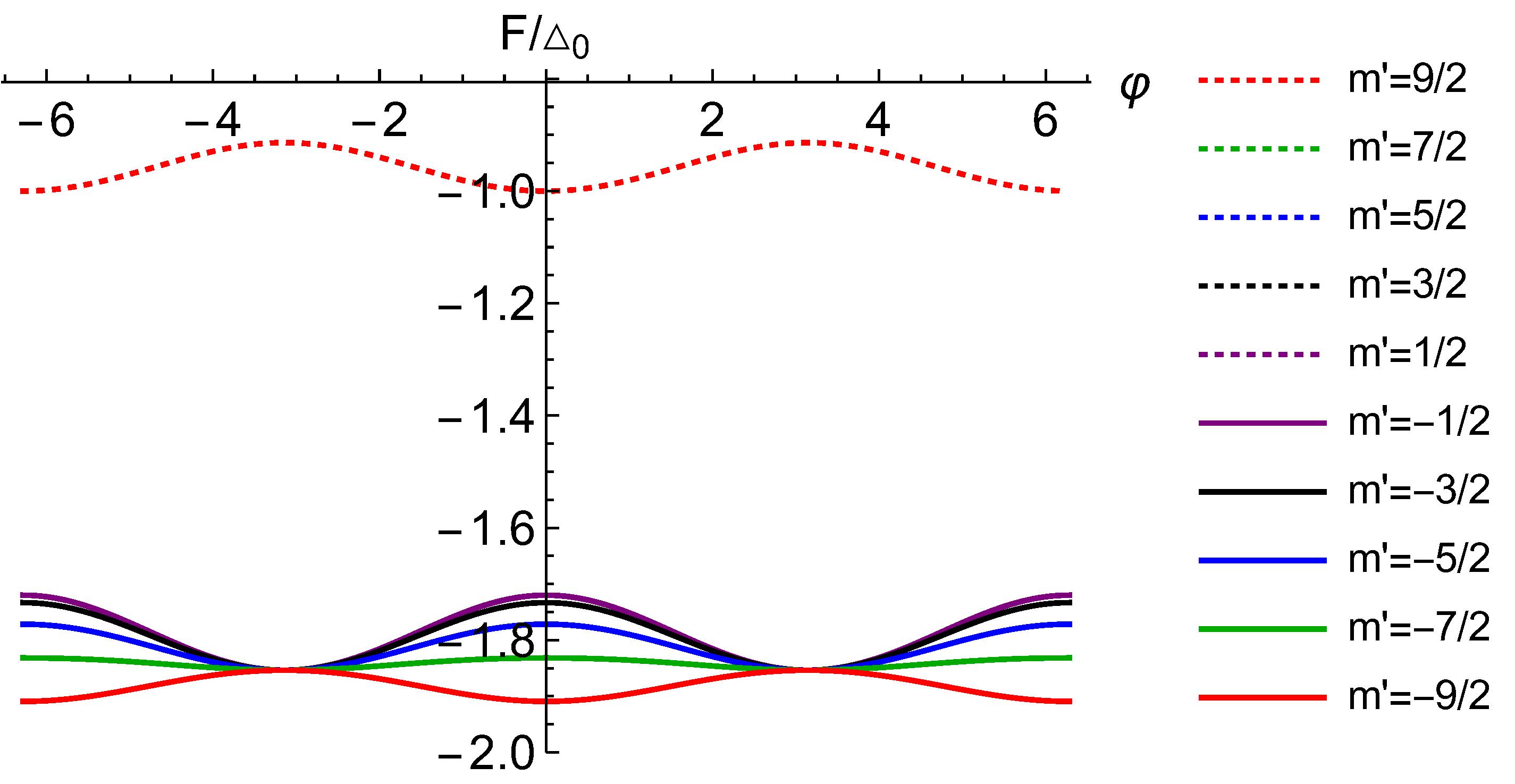}
\caption{\small \sl Free energy as a function of phase difference ($\varphi$). Parameters are $\Delta_{0}=1 meV$, $T/T_{c}=0.01$, $S=9/2$, $m'=\pm9/2,\pm7/2,\pm5/2,\pm3/2,\pm1/2$, $J=1$, $Z=0$. Free energy for $m'=7/2$ and $m'=-9/2$ are same and similarly for $m'=5/2$ and $m'=-7/2$, $m'=3/2$ and $m'=-5/2$, $m'=1/2$ and $m'=-3/2$ are same.}
\label{IB3}
\end{figure}

\begin{figure}[h]  
\includegraphics[width=.65\textwidth]{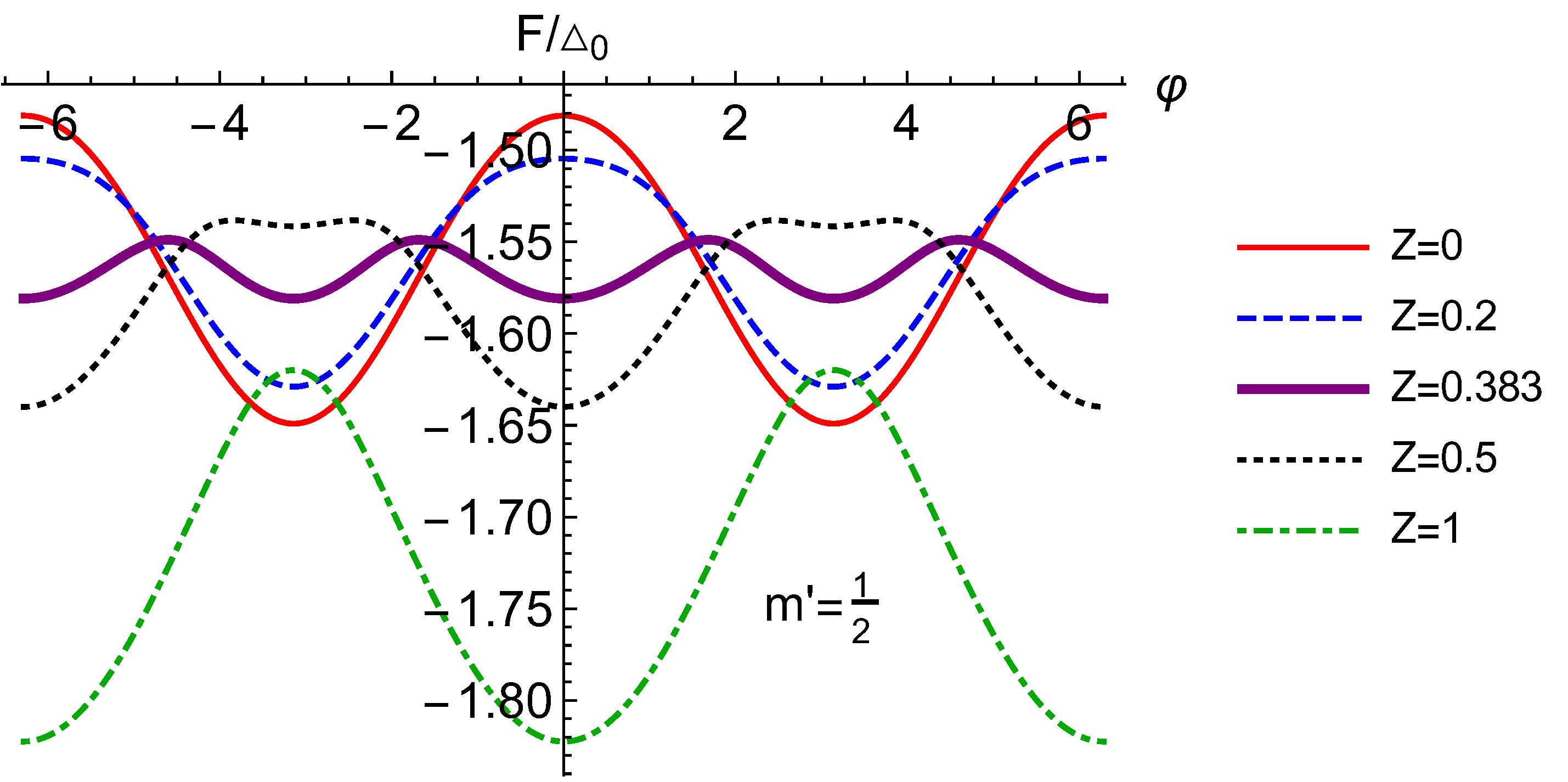}
\caption{\small \sl Free energy as a function of phase difference ($\varphi$) for different values of interface barrier strength ($Z$). Parameters are $\Delta_{0}=1 meV$,  $T/T_{c}=0.01$, $S=5/2$, $m'=1/2$, $J=1$.}
\label{FB}
\end{figure}

We can also determine the nature of the junction, i.e. $0$ or $\pi$ by the minimum of the free energy, which is given by 
\begin{equation}
F(\varphi)= -\frac{1}{\beta}\frac{1}{2\pi}\int_{0}^{2\pi}\ln\Big[\prod_{i}(1+e^{-\beta E_{i}(\varphi)})\Big]d(k_{F}a)= -\frac{2}{\beta}\frac{1}{2\pi}\int_{0}^{2\pi}\sum_{\sigma}\ln\Big[2\cosh\Big(\frac{\beta E_{\sigma}(\varphi)}{2}\Big)\Big]d(k_{F}a)
\end{equation}
In Fig.~\ref{IB3} we have plotted $F/\Delta_{0}$ as a function of phase difference for spin $S=9/2$ and different values of $m'$, we have considered a transparent junction ($Z=0$). In the same figure we see that the free energy for $m'=9/2$ is almost half than that of the other cases ($m'\neq 9/2$). A plausible reason for why these occurs could be that for $m'=9/2$ there is no spin flip process ($F_{2}=0$) while for the other cases $F_{2}$ ranges from $3$ to $5$. In Fig.~\ref{FB} we plot the Free energy for $S=5/2$ and $m'=1/2$ for different values of interface transparency $Z$. At particular value of $Z=0.383$ the Free energy shows a bistable behavior, i.e., the Free energy minima occurs at both $0$ and $\pi$ meaning that the ground state of the system does not occur at either $0$ or $\pi$ exclusively but is shared by both. These bistable junctions have a major role to play in quantum computation applications\cite{benj,lol,EE}.
\section{Effect of tunnel contacts}
\begin{figure}[h]  
\includegraphics[width=.99\textwidth]{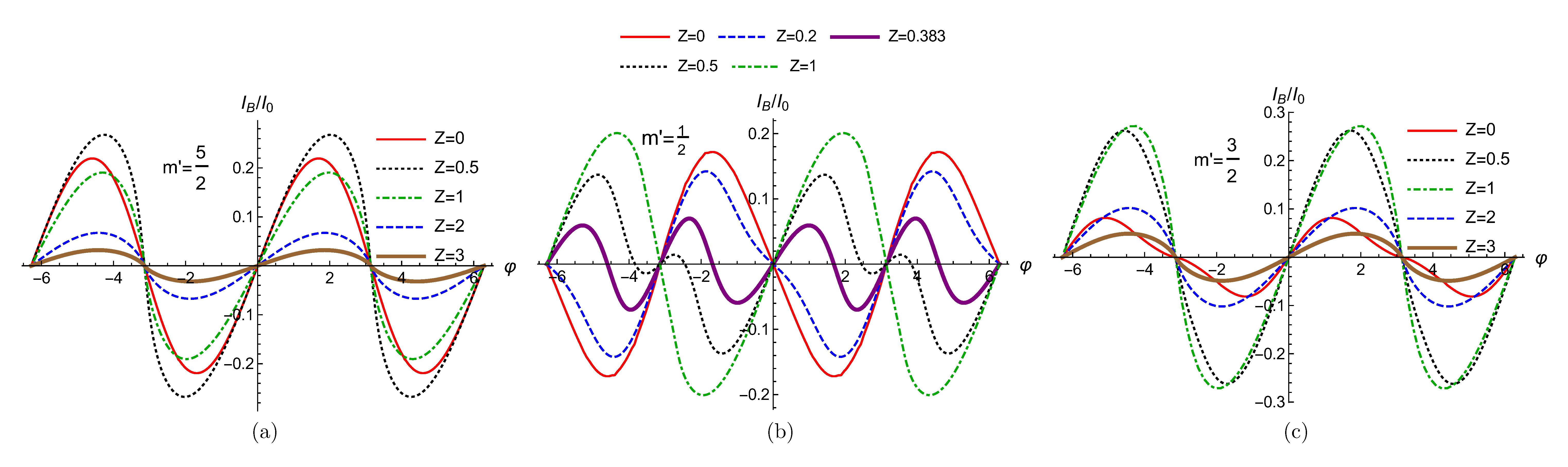}
\caption{\small \sl Josephson supercurrent as a function of phase difference ($\varphi$) for different values of interface barrier strength ($Z$). Parameters are $\Delta_{0}=1 meV$, $T/T_{c}=0.01$, $J=1$, $S=5/2$ and for (a) $m'=5/2$, (b) $m'=1/2$ and (c) $m'=3/2$. Josephson supercurrent for $m'=3/2$ and $m'=-5/2$ are same and similarly $m'=1/2$ and $m'=-3/2$ are same.}
\label{IB4}
\end{figure}

In Fig.~\ref{IB4} we plot the Josephson supercurrent as a function of phase difference for different values of interface barrier strength. From Fig.~\ref{IB4}(a) where $m'=5/2$ we see that there is no $\pi$ shift from transparent to tunnel regime and the ground state of the system always stays at $\varphi=0$. The reason that ground state stays at $\varphi=0$ in Fig.~\ref{IB4}(a) is because of the absence of spin flip processes as $S=m'=5/2$ and $F_{2}=0$. In Fig.~\ref{IB4}(b) the ground state of the system shifts from $\varphi=\pi$ to $\varphi=0$ as a function of $Z$. Infact for a transparent junction ($Z=0$) the ground state is at $\varphi=\pi$ and as we increase $Z$ we see the ground state shift from $\pi$ to $0$ state. Of course in this case as $S=5/2$ and $m'=1/2$ therefore the probability for the HSM to flip ($F_{2}\neq0$) is nonzero. Thus spin flip processes aid in the transition from $0$ to $\pi$ junction. Notably, this transition can be tunned by the transparency of the junction ($Z$) as is evident from Fig.~\ref{IB4}(b). Of course not all cases where in the HSM flips its spin leads to a transition from $0$ to $\pi$ state as is evident in Fig.~\ref{IB4}(c). In Fig.~\ref{IB4}(c) the ground state stays at $\varphi=0$, but here as $S=5/2$, $m'=3/2$ and $F_{2}\neq 0$, so spin flip processes occur in contrast to Fig.~\ref{IB4}(a). In Fig.~\ref{IB4} the strength of exchange interaction $J$ is taken as $1$. It has to be pointed out that $J$ has a nontrivial role in the $0$ to $\pi$ state transition as will be evident in the next section. Thus our conclusions regarding Fig.~\ref{IB4}(c) has to be qualified by the fact that we haven't focused on the issue of exchange interaction so far.
\section{Effect of exchange coupling}
\begin{figure}[h]  
\includegraphics[width=0.99\textwidth]{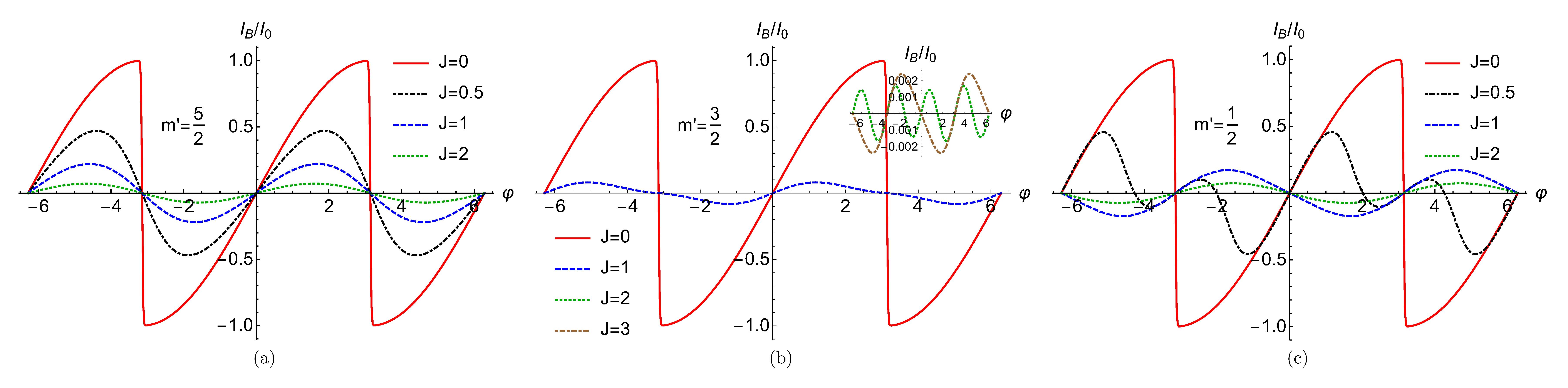}
\caption{\small \sl Josephson supercurrent as a function of phase difference ($\varphi$) for different values of exchange interaction ($J$). Parameters are $\Delta_{0}=1 meV$, $T/T_{c}=0.01$, $Z=0$, $S=5/2$ for (a) $m'=5/2$, (b) $m'=3/2$ and (c) $m'=1/2$. Josephson supercurrent for $m'=3/2$ and $m'=-5/2$ are same and similarly $m'=1/2$ and $m'=-3/2$ are same.}
\label{IB5}
\end{figure}

In Hamiltonian $H$, in Eq. (2) the term $J_{0}\delta(x)\vec s.\vec S$ represents the exchange coupling of strength $J_{0}$ between the electron with spin $\vec s$ and a HSM with spin $\vec S$.
In Fig.~\ref{IB5} the Josephson supercurrent is plotted as a function of phase difference for different values of strength of exchange interaction in the transparent regime. We choose $S=5/2$ and allowed values of $m'$. One sees for the no spin flip case there is no transition from $0$ to $\pi$ junction while for cases with spin flip one can see a $0$ to $\pi$ state transition. Thus all spin flip process i.e., $F_{2}\neq 0$ and with $J>2$ show $\pi$ junction behavior.
\section{Effect of electron-electron interaction (phenomenological):}
\begin{figure}[h]
\includegraphics[width=0.82\textwidth]{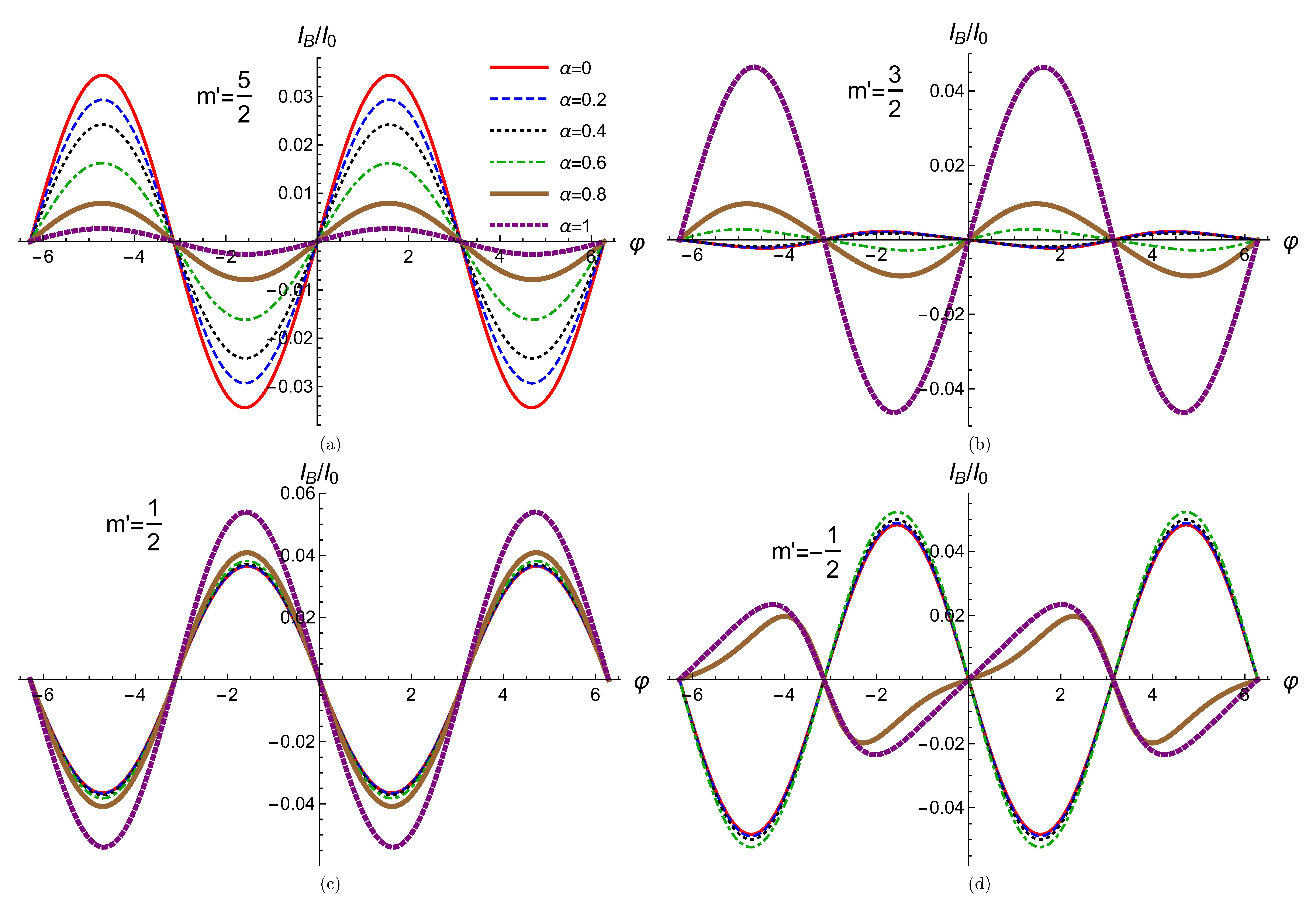}
\caption{\small \sl Josephson supercurrent as a function of phase difference ($\varphi$) for different values of electron-electron interaction strength ($\alpha$). Parameters are $\Delta_{0}=1 meV$, $D_{0}=100\Delta_{0}$, $T/T_{c}=0.01$, $J=3$, $Z=0.1$, $S=5/2$ and for (a) $m'=5/2$, (b) $m'=3/2$, (c) $m'=1/2$ and (d) $m'=-1/2$. Josephson supercurrent for $m'=3/2$ and $m'=-5/2$ are same and similarly $m'=1/2$ and $m'=-3/2$ are same.}
\label{IB6}
\end{figure}

We have considered a phenomenological\cite{Man,Ben} approach to electron-electron interactions. The effect of such interactions are included through an energy dependent transmission probability which is given as-
\begin{equation}
 T(E)=\frac{T_{0}\Big|\frac{E}{D_{0}}\Big|^{\alpha}}{1-T_{0}\Big(1-\Big|\frac{E}{D_{0}}\Big|^{\alpha}\Big)}
\end{equation}
with $T_{0}$ being the transparency of the metal superconductor interface in the absence of electron-electron interactions. $\alpha$ ($0<\alpha<1$) represents the electron-electron interaction strength ($\alpha=0$ corresponds to no interactions while $\alpha=1$ corresponds to a maximally interacting system), $D_{0}$ is a high energy cutoff obtained by the energy bandwidth of the electronic states.
Now for non-interacting case the parameter $Z$ is a constant and is related to the transmission probability $T_{0}$ as
\begin{equation}
 Z^{2}=\frac{1-T_{0}}{T_{0}}
\end{equation}
Now in presence of electron-electron interaction, $T_{0}$ is replaced by $T(E)$ in the above equation. Thus, the interface transparency $Z$ which is considered identical at both interfaces will be energy dependent and will change from $Z$ to $Z_{ee}$:
\begin{equation}
Z_{ee}^2=\Big|\frac{E}{D_{0}}\Big|^{-\alpha}\frac{1-T_{0}}{T_{0}}=\Big|\frac{E}{D_{0}}\Big|^{-\alpha}Z^2 
\end{equation}
For $Z=0$ $(T_{0}=1)$, $Z_{ee}=0$ which implies that for a transparent interface electron-electron interaction have no effect on electronic transport.
In Fig.~\ref{IB6} we plot the Josephson supercurrent as a function of phase difference for different values of electron-electron interaction  parameter $\alpha$. We see that for $m'=5/2,1/2,-3/2$ there is no $0$-$\pi$ transition with increase of electron-electron interaction strength. But for $m'=3/2,-1/2,-5/2$ there is a change from $\pi$ to $0$ junction with increase of electron-electron interaction strength ($\alpha$).
\section{Effect of high spin/magnetic moment states}
\begin{figure}[h]
\includegraphics[width=0.85\textwidth]{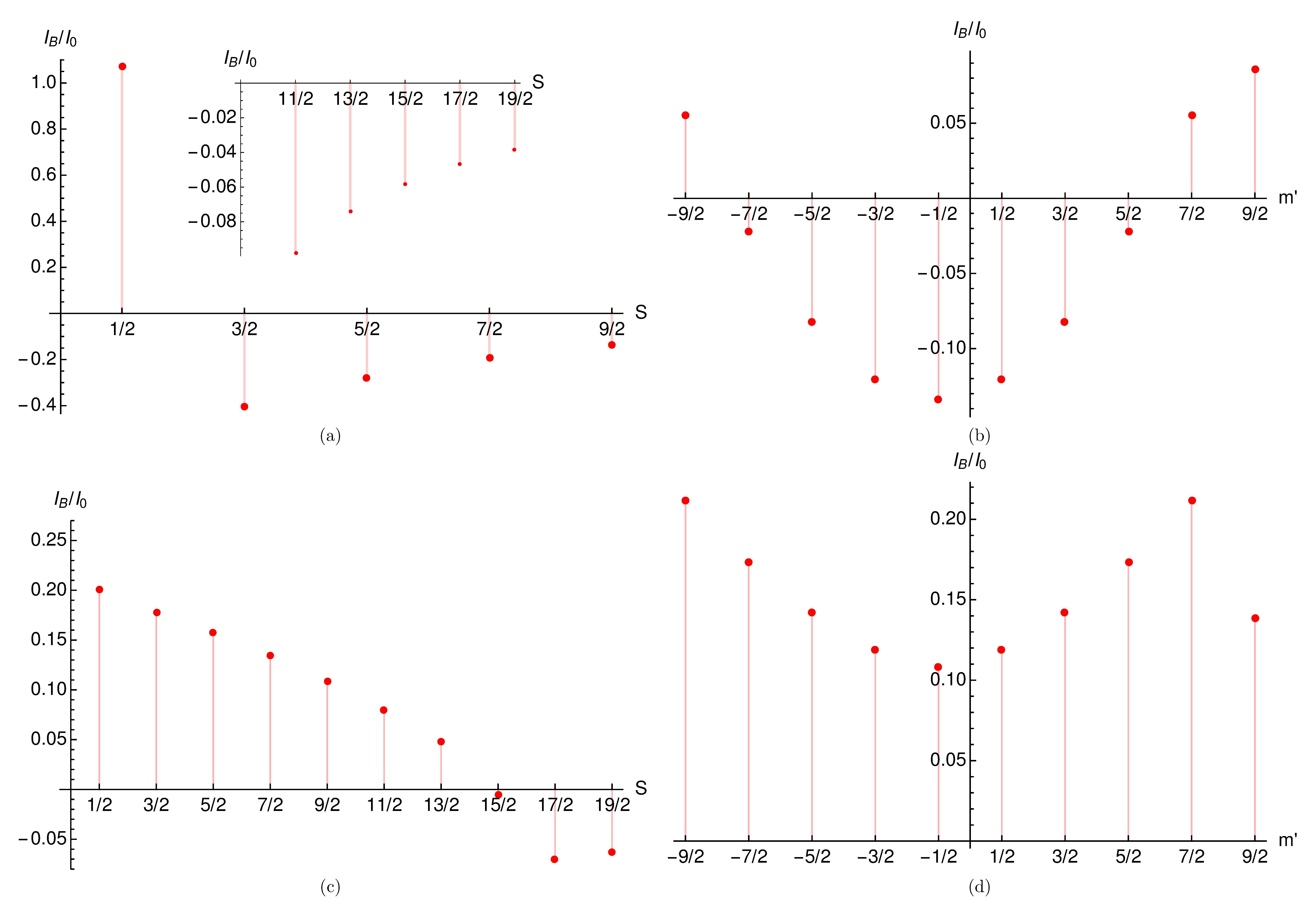}
\caption{\small \sl  (a) Josephson supercurrent vs HSM spin ($S$). Parameters are $\Delta_{0}=1 meV$, $T/T_{c}=0.01$, $\varphi=\pi/2$, $J=1$, $m'=-1/2$, $Z=0$, (b) Josephson supercurrent vs HSM magnetic moment ($m'$). Parameters are $\Delta_{0}=1 meV$, $T/T_{c}=0.01$, $\varphi=\pi/2$, $J=1$, $S=9/2$, $Z=0$, (c) Josephson supercurrent vs HSM spin ($S$). Parameters are $\Delta_{0}=1 meV$, $T/T_{c}=0.01$, $\varphi=\pi/2$, $J=1$, $m'=-1/2$, $Z=1$, (d) Josephson supercurrent vs HSM magnetic moment ($m'$). Parameters are $\Delta_{0}=1 meV$, $T/T_{c}=0.01$, $\varphi=\pi/2$, $J=1$, $S=9/2$, $Z=1$.}
\label{IB7}
\end{figure}

Since we have a high spin magnetic impurity(HSM) it is imperative for us to study high spin states of our HSM. In Fig.~\ref{IB7}(a) we see that Josephson supercurrent at $\varphi=\pi/2$ is positive for $S=1/2$, but as we increase spin ($S$) of HSM it changes to negative from $S=3/2$ to $S=9/2$. We choose phase difference $\varphi=\pi/2$ to see the sign change of the Josephson supercurrent. In the inset of Fig.~\ref{IB7}(a) we plot the Josephson supercurrent for still higher spin states of HSM ($S=11/2-19/2$). In Fig.~\ref{IB7}(a) for all different values of $S$ the magnetic moment of HSM $m'=-1/2$ and the junction transparency $Z=0$. The reason for the change in sign in the Josephson supercurrent can be guessed from the fact that the spin flip probability ($F_{2}$) of the HSM for negative Josephson supercurrent is greater than $1$. This previous statement is however subject to qualification-negative supercurrent for low spin states of HSM require smaller values of spin flip probability $F_{2}$ than do high spin states of HSM. In Fig.~\ref{IB7}(b) we look at the effect of spin magnetic moment states on Josephson supercurrent. We consider the spin $S$ of HSM to be $9/2$. The Josephson supercurrent changes sign with $m'$. One can clearly see when the spin flip probability of HSM i.e., $F_{2}>3$ the Josephson supercurrent is negative but for flip probability $F_{2}<3$ the Josephson supercurrent is positive for a transparent junction $Z=0$. We see in Fig.~\ref{IB7}(c) the possibility of a $\pi$ junction also at $Z=1$ (intermediate transparency). In Fig.~\ref{IB7}(c) we plot the Josephson supercurrent including still higher spin states of HSM ($S=1/2-19/2$). In {supplementary material section IV} we juxtapose the spin state $S$, magnetic moment $m'$ and spin flip probability $F_{2}$ of HSM in a tabular format. Finally in Fig.~\ref{IB7}(d) we plot the Josephson supercurrent at $Z=1$ (non transparent junction) as a function of spin magnetic moment $m'$ for $S=9/2$. We see non transparent junction inhibit a $0-\pi$ junction transition for $S=9/2$. However, one has to qualify the aforesaid statement by looking at Fig.~\ref{IB7}(c). In Fig.~\ref{IB7}(c) we see that a finite $Z$ (equal $1$) can act as a barrier to the $0-\pi$ junction transition. To overcome this barrier one needs to go to still higher spin states like $S=15/2-19/2$. Thus in Fig.~\ref{IB7}(d) instead of plotting for $S=9/2$ if we had plotted for $S=15/2-19/2$ we would have seen a $0-\pi$ junction transition for some value of $m'$. So to conclude this section for transparent interfaces spin flip processes lead to a $0$ to $\pi$ junction transition. However, when junction transparency reduces one has to go to much higher spin states to see a $0-\pi$ junction transition. The moral of the story is a finite $Z$ inhibits $0-\pi$ transition but a large $S$ can overcome the $Z$ barrier. { The $\pi-$shift seen due to change in $S$ can be experimentally implemented. One can control the impurity spin $S$ optically as shown in Refs.\cite{gov,GOV}.}
\section{Experimental realization and Conclusions}
In this paper we have provided an exhaustive study of the nature of the $0$ to $\pi$ Josephson junction transition in presence of a high spin magnetic impurity(HSM). We have studied various aspects of the problem like the strength of the exchange interaction ($J$) between HSM and charge carriers (Section 9), the effect of electron-electron interactions ($\alpha$) albeit phenomenologically (Section 10), effect of junction transparencies ($Z$) (Section 8) and of course the high spin states $S$, spin magnetic moment $m'$ of the HSM itself (Section 11). We identify spin flip probability of the HSM as the key to understand the $0$ to $\pi$ junction transition. We also focused on applications of our junction in quantum computation proposals (Section 7). {The set-up as shown in Figure 1 can be easily realized in the lab. Superconductor-Normal metal-Superconductor Josephson junctions have been experimentally realized since long\cite{and}. High spin magnetic impurities have been realized since 20 years\cite{tku}. The amalgamation of a Superconductor-Normal metal-Superconductor (SNS) junction with a high spin magnetic impurity shouldn't be difficult, especially with a s-wave superconductor like Aluminum or Lead it should be perfectly possible. $\pi$ Josephson junction with a quantum dot sandwiched between two superconductors has been demonstrated experimentally in Ref.~\cite{cleu}. They observe a gate-controlled transition from the $0$ to the $\pi$ state. Further, in Ref.~\cite{peng} they look at the Josephson effect in a quantum spin Hall system coupled with a localized magnetic impurity.}
Our work will help experimentalists in designing tunable $\pi$ junctions without taking recourse to Ferromagnets or high $T_{c}$ superconductors or any applied magnetic fields but with only a magnetic impurity.

\section*{Acknowledgments}
This work was supported by the grant ``Non-local correlations in nanoscale systems: Role of decoherence, interactions, disorder and pairing symmetry'' from SCIENCE \& ENGINEERING RESEARCH BOARD, New Delhi, Government of India, Grant No.  EMR/2015/001836,
Principal Investigator: Dr. Colin Benjamin, National Institute of Science Education and Research, Bhubaneswar, India.
\section*{Author contributions statement}
C.B. conceived the proposal,  S.P. did the calculations on the advice of C.B., C.B. and S.P. analyzed the results and wrote the paper.  Both  authors reviewed the manuscript. 
\section*{Competing interests statement}
The authors have no competing  interests.
\newpage
{\hskip 2.0in \Large \bf\center Supplementary Material:}
In section 13 we first introduce our model, provide a theoretical background to our study with wavefunctions and boundary conditions to calculate the Josephson current. In section 14 we give the explicit form of $8\times8$ matrix $M$. The explicit form of Andreev bound states is given in section 15. Finally, in section 16 we supply spin flip probability ($F_{2}$) values of the high spin magnetic impurity (HSM) for different impurity spin ($S$) and magnetic moment ($m'$) in a tabular format.
\section{Wavefunctions and boundary conditions in the Josephson junction in presence of a high spin magnetic impurity}
We consider a system consists of two normal metals with a HSM sandwiched between two conventional s-wave singlet superconductors. Our model is shown in Fig.~1, it depicts a HSM at $x=0$ and two superconductors at $x<-a/2$ and $x>a/2$. There are normal metal regions in $-a/2<x<0$ and $0<x<a/2$.
\subsection{Wavefunctions}
There can be eight different types of quasiparticle injection into our system: an electron-like quasiparticle (ELQ) with spin up or down or a hole-like quasiparticle (HLQ) with spin up or down injected from either the left or from the right superconducting electrode. For the injection of spin up electron in left superconductor, the wave function is given by\cite{LINDER}-
\begin{eqnarray}
\psi_{SL}(x)=\begin{pmatrix}u\\
                              0\\
                              0\\ 
                              v
                              \end{pmatrix}e^{iq_{+}x}\phi_{m'}^{S}+r_{ee}^{\uparrow\uparrow}\begin{pmatrix}
                              u\\
                              0\\
                              0\\
                              v
                             \end{pmatrix}e^{-iq_{+}x}\phi_{m'}^{S}+r_{ee}^{\uparrow\downarrow}\begin{pmatrix}
                             0\\
                             u\\
                            -v\\
                             0
                             \end{pmatrix}e^{-iq_{+}x}\phi_{m'+1}^{S}+r_{eh}^{\uparrow\uparrow}\begin{pmatrix}
                             0\\
                            -v\\
                             u\\
                             0
                             \end{pmatrix}e^{iq_{-}x}\phi_{m'+1}^{S}+r_{eh}^{\uparrow\downarrow}\begin{pmatrix}
                             v\\
                             0\\
                             0\\
                             u
                             \end{pmatrix}e^{iq_{-}x}\phi_{m'}^{S},\nonumber\\\mbox{for $x<-\frac{a}{2}$}\nonumber
                             \end{eqnarray}
The amplitudes $r_{ee}^{\uparrow\uparrow},r_{ee}^{\uparrow\downarrow},r_{eh}^{\uparrow\uparrow},r_{eh}^{\uparrow\downarrow}$ represent normal reflection, normal reflection with spin flip, Andreev reflection with spin flip and Andreev reflection without flip respectively.\\
The corresponding wave function in the right superconductor is-
\[\psi_{SR}(x)=t_{ee}^{\uparrow\uparrow}\begin{pmatrix}
                              ue^{i\varphi}\\
                              0\\
                              0\\
                              v
                             \end{pmatrix}e^{iq_{+}x}\phi_{m'}^{S}+t_{ee}^{\uparrow\downarrow}\begin{pmatrix}
                             0\\
                             ue^{i\varphi}\\
                             -v\\
                             0
                             \end{pmatrix}e^{iq_{+}x}\phi_{m'+1}^{S}+t_{eh}^{\uparrow\uparrow}\begin{pmatrix}
                             0\\
                             -ve^{i\varphi}\\
                             u\\
                             0
                             \end{pmatrix}e^{-iq_{-}x}\phi_{m'+1}^{S}+t_{eh}^{\uparrow\downarrow}\begin{pmatrix}
                             ve^{i\varphi}\\
                             0\\
                             0\\
                             u
                             \end{pmatrix}e^{-iq_{-}x}\phi_{m'}^{S},\mbox{for $x>\frac{a}{2}$}\]
where $t_{ee}^{\uparrow\uparrow},t_{ee}^{\uparrow\downarrow},t_{eh}^{\uparrow\uparrow},t_{eh}^{\uparrow\downarrow}$ are the transmission amplitudes, corresponding to the reflection process described above and $\varphi=\varphi_{R}-\varphi_{L}$ is the phase difference between right side and left side superconductor. $\phi_{m'}^{S}$ is the eigenspinor of the HSM, with its $S^{z}$ operator acting as-
$S^{z}\phi_{m'}^{S} = m'\phi_{m'}^{S}$, with $m'$ being the spin magnetic moment of the HSM. The BCS coherence factors are defined as 
$u=\sqrt{\frac{1}{2}\Bigg(1+\frac{\sqrt{E^2-\Delta_{0}^{2}}}{E}\Bigg)}$, $v=\sqrt{\frac{1}{2}\Bigg(1-\frac{\sqrt{E^2-\Delta_{0}^{2}}}{E}\Bigg)}$.
$q_{\pm}=\sqrt{\frac{2m^{\star}}{\hbar^2}(E_{F}\pm \sqrt{E^2-\Delta_{0}^2})}$ is the wavevector for electron-like quasiparticle ($q_{+}$) and hole-like quasiparticle ($q_{-}$) in the left and right superconducting wavefunctions, $\psi_{SL}$ and $\psi_{SR}$.
The wavefunction in the normal metal region ($N_{1}$) is given by-
\begin{eqnarray}
\psi_{N_{1}}(x)=(ee^{ik_{e}(x+a/2)}+fe^{-ik_{e}x})\begin{pmatrix}
                                                     1\\
                                                     0\\
                                                     0\\
                                                     0
                                                    \end{pmatrix}\phi_{m'}^{S}+(e^{\prime} e^{ik_{e}(x+a/2)}+f^{\prime} e^{-ik_{e}x})\begin{pmatrix}
                                                     0\\
                                                     1\\
                                                     0\\
                                                     0
                                                     \end{pmatrix}\phi_{m'+1}^{S}\nonumber\\+(ge^{-ik_{h}(x+a/2)}+he^{ik_{h}x})\begin{pmatrix}
                                                     0\\
                                                     0\\
                                                     1\\
                                                     0
                                                     \end{pmatrix}\phi_{m'+1}^{S}+(g^{\prime} e^{-ik_{h}(x+a/2)}+h^{\prime} e^{ik_{h}x})\begin{pmatrix}
                                                     0\\
                                                     0\\
                                                     0\\
                                                     1
                                                     \end{pmatrix}\phi_{m'}^{S},\mbox{for $-\frac{a}{2}<x<0$}\nonumber
                                                    \end{eqnarray}
Similarly the wavefunction in the normal metal region ($N_{2}$) is given by-
\begin{eqnarray}
\psi_{N_{2}}(x)=(a_{0}e^{ik_{e}x}+be^{-ik_{e}(x-a/2)})\begin{pmatrix}
                                                     1\\
                                                     0\\
                                                     0\\
                                                     0
                                                    \end{pmatrix}\phi_{m'}^{S}+(a^{\prime} e^{ik_{e}x}+b^{\prime} e^{-ik_{e}(x-a/2)})\begin{pmatrix}
                                                     0\\
                                                     1\\
                                                     0\\
                                                     0
                                                     \end{pmatrix}\phi_{m'+1}^{S}\nonumber\\+(ce^{-ik_{h}x}+de^{ik_{h}(x-a/2)})\begin{pmatrix}
                                                     0\\
                                                     0\\
                                                     1\\
                                                     0
                                                     \end{pmatrix}\phi_{m'+1}^{S}+(c^{\prime} e^{-ik_{h}x}+d^{\prime} e^{ik_{h}(x-a/2)})\begin{pmatrix}
                                                     0\\
                                                     0\\
                                                     0\\
                                                     1
                                                     \end{pmatrix}\phi_{m'}^{S},\mbox{for $0<x<\frac{a}{2}$}\nonumber
                                                    \end{eqnarray}
$k_{e,h}=\sqrt{\frac{2m^{\star}}{\hbar^2}(E_{F}\pm E)}$ is the wave vector in the normal metals.
In our work we have used the Andreev approximation\cite{Kri} $q_{+}=q_{-}=k_{F}$ and $k_{e,h}\approx k_{F}\pm \frac{k_{F}E}{2E_{F}}$, where $k_{F}$ is the Fermi wavevector, with $E_{F}>>\Delta$.\\
\subsection{Boundary conditions}
The boundary conditions at $x=-a/2:$\\
\begin{equation}
 \psi_{SL}(x)=\psi_{N_{1}}(x),\mbox{(continuity of wavefunctions)}\nonumber
\end{equation}
\begin{equation}
\frac{d\psi_{N_{1}}}{dx}-\frac{d\psi_{SL}}{dx}=\frac{2m^{\star}V}{\hbar^2}\psi_{N_{1}},\mbox{(discontinuity in first derivative)}\nonumber 
\end{equation}
and at $x=0:$
\begin{equation}
\psi_{N_{1}}(x)=\psi_{N_{2}}(x)\nonumber 
\end{equation}
\begin{equation}
\frac{d\psi_{N_{2}}}{dx}-\frac{d\psi_{N_{1}}}{dx}=-\frac{2m^{\star}J_{0}\vec s.\vec S}{\hbar^2} \psi_{N_{1}}\nonumber 
\end{equation}
where $\vec s.\vec S$ is the exchange operator in the Hamiltonian and is given by $\vec s.\vec S=s^{z}S^{z}+\frac{1}{2}(s^{-}S^{+}+s^{+}S^{-})$;
\begin{equation}
\vec s.\vec S\begin{pmatrix}
               1\\
               0\\
               0\\
               0
              \end{pmatrix}\phi_{m'}^{S}=mm'\begin{pmatrix}
              1\\
              0\\
              0\\
              0
             \end{pmatrix}\phi_{m'}^{S}+\frac{1}{2}F_{1}F_{2}\begin{pmatrix}
             0\\
             1\\
             0\\
             0
             \end{pmatrix}\phi_{m'+1}^{S}\nonumber
\end{equation}
and \begin{equation} 
\vec s.\vec S\begin{pmatrix}
              0\\
              1\\
              0\\
              0\\
             \end{pmatrix}\phi_{m'+1}^{S}=(m-1)(m'+1)\begin{pmatrix}
             0\\
             1\\
             0\\
             0
             \end{pmatrix}\phi_{m'+1}^{S}+\frac{1}{2}F_{1}F_{2}\begin{pmatrix}
                                                                1\\
                                                                0\\
                                                                0\\
                                                                0
                                                               \end{pmatrix}\phi_{m'}^{S}\nonumber
                                                               \end{equation}
Here $F_{1}=\sqrt{(s+m)(s-m+1)}$ is the spin-flip probability for electron, $F_{2}=\sqrt{(S-m')(S+m'+1)}$ is the spin-flip probability\cite{AJP} for HSM, $m$ is the spin magnetic moment of the spin up electron ($m=1/2$) and $m-1$ is the spin magnetic moment of the spin down electron ($m-1=-1/2$). $s^{\pm} = s_{x}\pm is_{y}$ and $S^{\pm} = S_{x}\pm iS_{y}$ are the raising and lowering spin operators.\\ 
Finally, at $x=a/2:$
\begin{equation}
\psi_{N_{2}}(x)=\psi_{SR}(x)\nonumber 
\end{equation}
\begin{equation}
\frac{d\psi_{SR}}{dx}-\frac{d\psi_{N_{2}}}{dx}=\frac{2m^{\star}V}{\hbar^2}\psi_{N_{2}}.\nonumber
\end{equation}
We will later use the dimensionless parameter  $J=\frac{m^{\star}J_{0}}{\hbar^2k_{F}}$ as a measure of strength of exchange interaction and $Z = \frac{V}{V_{0}}$, with $V_{0}=\frac{\hbar^2k_{F}}{m^{*}}$ as a measure of interface transparency. Thus a value of $Z=5$ (say) means interface potential $V=5V_{0}$, with $V$ in units of $V_{0}$. By using above boundary conditions one can get the different scattering amplitudes. The wave functions for the other seven types of quasiparticle injection process are constructed in the same way.
\section{Explicit form of Matrix M}
To calculate the bound state contribution of total Josephson current we introduce a $8\times8$ matrix $M$ in Eq.~(4) in section 3.2 of our paper
which is given by-
\[
M=\begin{pmatrix}
   M_{11} & M_{12} & M_{13} & M_{14} & M_{15} & M_{16} & M_{17} & M_{18}\\
   M_{21} & M_{22} & M_{23} & M_{24} & M_{25} & M_{26} & M_{27} & M_{28}\\
   M_{31} & M_{32} & M_{33} & M_{34} & M_{35} & M_{36} & M_{37} & M_{38}\\
   M_{41} & M_{42} & M_{43} & M_{44} & M_{45} & M_{46} & M_{47} & M_{48}\\
   M_{51} & M_{52} & M_{53} & M_{54} & M_{55} & M_{56} & M_{57} & M_{58}\\
   M_{61} & M_{62} & M_{63} & M_{64} & M_{65} & M_{66} & M_{67} & M_{68}\\
   M_{71} & M_{72} & M_{73} & M_{74} & M_{75} & M_{76} & M_{77} & M_{78}\\
   M_{81} & M_{82} & M_{83} & M_{84} & M_{85} & M_{86} & M_{87} & M_{88}\\
\end{pmatrix}\]
where,
\begin{align}
\begin{split}
 M_{11}={}& (-ie^{ik_{F}a}u+e^{ik_{F}a}uZ-e^{2ik_{F}a}uZ)\\
 M_{12}={}& 0\\
 M_{13}={}& 0\\ 
 M_{14}={}& -v(ie^{ik_{F}a}-Z+e^{ik_{F}a}Z)\\ 
 M_{15}={}& (ie^{ik_{F}a+i\varphi}u-e^{ik_{F}a+i\varphi}uZ+e^{2ik_{F}a+i\varphi}uZ)\\ 
 M_{16}={}& 0\\ 
 M_{17}={}& 0\\
 M_{18}={}& (ie^{ik_{F}a+i\varphi}v+e^{ik_{F}a+i\varphi}vZ-e^{i\varphi}vZ)\\
 M_{21}={}& 0\\
 M_{22}={}& (i e^{ik_{F}a}u-e^{ik_{F}a}uZ+e^{2ik_{F}a}uZ)\\
 M_{23}={}& -v(ie^{ik_{F}a}-Z+e^{ik_{F}a}Z)\\
 M_{24}={}& 0\\
 M_{25}={}& 0\\
 M_{26}={}& (-ie^{ik_{F}a+i\varphi}u+e^{ik_{F}a+i\varphi}uZ-e^{2ik_{F}a+i\varphi}uZ)\\ 
 M_{27}={}& (ie^{ik_{F}a+i\varphi}v+e^{ik_{F}a+i\varphi}vZ-e^{i\varphi}vZ)\\
 M_{28}={}& 0\\
 M_{31}={}& 0\\
 M_{32}={}& (ie^{ik_{F}a} v-e^{ik_{F}a}vZ+e^{2ik_{F}a}vZ)\\
 M_{33}={}& -u(ie^{ik_{F}a}-Z+e^{ik_{F}a}Z)\\
 M_{34}={}& 0\\
 M_{35}={}& 0\\
 M_{36}={}& (-ie^{ik_{F}a}v+e^{ik_{F}a}vZ-e^{2ik_{F}a}vZ)\\ 
 M_{37}={}& (ie^{ik_{F}a}u-uZ+e^{ik_{F}a}uZ)\\ 
 M_{38}={}& 0\\
 M_{41}={}& (-ie^{ik_{F}a}v+e^{ik_{F}a}vZ-e^{2ik_{F}a}vZ)\\ 
 M_{42}={}& 0\\ 
 M_{43}={}& 0\\
 M_{44}={}& -u(ie^{ik_{F}a}-Z+e^{ik_{F}a}Z)\\
 M_{45}={}& (ie^{ik_{F}a}v-e^{ik_{F}a}vZ+e^{2ik_{F}a}vZ)\\ 
 M_{46}={}& 0\\ 
 M_{47}={}& 0\\
 M_{48}={}& (ie^{ik_{F}a}u-uZ+e^{ik_{F}a}uZ)\\
 M_{51}={}& (2e^{2ik_{F}a}Jmm'u(i-2Z)+2e^{3ik_{F}a}(-i+Jmm')uZ+2e^{ik_{F}a}(i+Jmm')u(-i+Z))\\ 
 M_{52}={}& (e^{2ik_{F}a}F_{1}F_{2}Ju(i-2Z)+e^{3ik_{F}a}F_{1}F_{2}JuZ+e^{ik_{F}a}F_{1}F_{2}Ju(-i+Z))\\ 
 M_{53}={}& (-F_{1}F_{2}JvZ-e^{2ik_{F}a}F_{1}F_{2}Jv(i+Z)+e^{ik_{F}a}F_{1}F_{2}Jv(i+2Z))\\
 M_{54}={}& (2e^{2ik_{F}a}v(1-iZ)+2(i+Jmm')vZ+2e^{2ik_{F}a}Jmm'v(i+Z)-2e^{ik_{F}a}Jmm'v(i+2Z))\\
 M_{55}={}& -2e^{2ik_{F}a+i\varphi}u\\ 
 M_{56}={}& 0\\ 
 M_{57}={}& 0\\
 M_{58}={}& -2e^{i(k_{F}a+\varphi)}v\\
 M_{61}={}& (e^{2ik_{F}a}F_{1}F_{2}Ju(i-2Z)+e^{3ik_{F}a}F_{1}F_{2}JuZ+e^{ik_{F}a}F_{1}F_{2}Ju(-i+Z))\\ 
 M_{62}={}& 2e^{ik_{F}a}u(1+(-1+e^{ik_{F}a})J(-1+m)(1+m')(i+(-1+e^{ik_{F}a})Z)+2e^{ik_{F}a}Z\sin(k_{F}a))\nonumber
 \end{split}
 \end{align}
 \begin{align}
 \begin{split}
 M_{63}={}& 2v((-i-J(-1+m)(1+m'))Z-e^{2ik_{F}a}(-i+J(-1+m)(1+m'))(i+Z)+e^{ik_{F}a}J(-1+m)(1+m')(i+2Z))\\ 
 M_{64}={}& (F_{1}F_{2}JvZ+e^{2ik_{F}a}F_{1}F_{2}Jv(i+Z)-e^{ik_{F}a}F_{1}F_{2}Jv(i+2Z))\\
 M_{65}={}& 0\\ 
 M_{66}={}& -2e^{ik_{F}a+i(k_{F}a+\varphi)}u\\ 
 M_{67}={}& 2e^{i(k_{F}a+\varphi)}v\\ 
 M_{68}={}& 0\\
 M_{71}={}& e^{-ik_{F}a}(e^{2ik_{F}a}F_{1}F_{2}Jv(i-2Z)+e^{3ik_{F}a}F_{1}F_{2}JvZ+e^{ik_{F}a}F_{1}F_{2}Jv(-i+Z))\\
 M_{72}={}& 2v(-(-1+e^{ik_{F}a})J(-1+m)(1+m')(i+(-1+e^{ik_{F}a})Z)+i(i+(-1+e^{2ik_{F}a})Z))\\ 
 M_{73}={}& e^{-ik_{F}a}(2(i+J(m-1)(m'+1))uZ+2e^{2ik_{F}a}(-i+J(m-1)(m'+1))u(i+ Z)-2e^{ik_{F}a}J(m-1)(m'+1)u(i+2Z))\\
 M_{74}={}& e^{-ik_{F}a}(F_{1}F_{2}JuZ+e^{2ik_{F}a}F_{1}F_{2}Ju(i+Z)-e^{ik_{F}a}F_{1}F_{2}Ju(i+2Z))\\ 
 M_{75}={}& 0\\
 M_{76}={}& 2e^{ik_{F}a}v\\ 
 M_{77}={}& -2u\\ 
 M_{78}={}& 0\\
 M_{81}={}& e^{-ik_{F}a}(2e^{ik_{F}a}v(1+iZ)+2e^{3ik_{F}a}(-i+Jmm')vZ+2e^{ik_{F}a}Jmm'v(-i+Z)-2e^{2ik_{F}a}Jmm'v(-i+2Z))\\
 M_{82}={}& e^{-ik_{F}a}(-e^{2ik_{F}a}F_{1}F_{2}Jv(i-2Z)-e^{3ik_{F}a}F_{1}F_{2}JvZ-e^{ik_{F}a}F_{1}F_{2}Jv(-i+ Z))\\\nonumber
 M_{83}={}& e^{-ik_{F}a}(F_{1}F_{2}JuZ+e^{2ik_{F}a}F_{1}F_{2}Ju(i+Z)-e^{ik_{F}a}F_{1}F_{2}Ju(i+2Z))\\ 
 M_{84}={}& e^{-ik_{F}a}(2(i+Jmm')uZ+2e^{2i k_{F}a}(-i+Jmm')u(i+Z)-2e^{ik_{F}a}Jmm'u(i+2Z))\\ 
 M_{85}={}& -2e^{ik_{F}a}v\\ 
 M_{86}={}& 0\\ 
 M_{87}={}& 0\\ 
 M_{88}={}& -2u\\\nonumber
\end{split}\nonumber
\end{align}
\section{Explicit form of Andreev bound states}
In Eq.~(5) Andreev bound state expression, we introduce the terms $A(\varphi)$, $B(\varphi)$ and $C$ in section 3.2 of our paper. The explicit form of $A(\varphi)$, $B(\varphi)$ and $C$ is given by
\begin{align}
\begin{split}
A(\varphi)={}& -2e^{3ik_{F}a}((1+2Z^2)(8(1+2Z^2)^2+J^4(F_{2}^2+m'+m'^2)^2(1+10(Z^2+Z^4))+J^2(3+6m'(1+m')+8Z^2\\
    & +2(F_{2}^2-8(F_{2}^2+m'+m'^2)Z^2-4(-1+2F_{2}^2+2m'(1+m'))Z^4)))+2Z^2(1+2Z^2)(16+16J^3(F_{2}^2+m'+m'^2)Z\\
    & -16Z^2+3J^4(F_{2}^2+m'+m'^2)^2(-1+Z^2)+4J^2(-1+2F_{2}^2+2m'(1+m'))(-1 + Z^2))\cos(2k_{F}a)-2Z^3(16Z(-3+Z^2)\\
    & +J^4(F_{2}^2+m'+m'^2)^2Z(-3+Z^2)+4J^2(-1+2F_{2}^2+2m'(1+m'))Z(-3+Z^2)+16J(-1+3Z^2)+4J^3(F_{2}^2+m'\\
    & +m'^2)(-1+3Z^2))\cos(3k_{F}a)+8\cos(\varphi)+(16Z^2-J^2(-1+2F_{2}^2-2m'(1+m'))(1+2Z^2))\cos(\varphi)+2Z\cos(k_{F}a)(8Z\\ 
    & +12J(1+2Z^2)^2+J^2Z(-3-2F_{2}^2-6m'(1+m')-4Z^2+8(F_{2}^2+m'+m'^2)Z^2+4(-1+2F_{2}^2+2m'(1+m'))Z^4)\\
    & +16(Z^3+Z^5)-4J^3(F_{2}^2+m'+m'^2)(1+5(Z^2+Z^4))-3J^4(F_{2}^2+m'+m'^2)^2Z(1+5(Z^2+Z^4))+(8Z+J(4\\
    & +J(-1+2F_{2}^2-2m'(1+m'))Z))\cos(\varphi))+2Z(-8+12JZ(1+2Z^2)^2-16(Z^2+Z^4)+J^2(3+2F_{2}^2+6m'(1+m')\\
    & +4Z^2-8(F_{2}^2+m'+m'^2)Z^2-4(-1+2F_{2}^2+2m'(1+m'))Z^4)+3J^4(F_{2}^2+m'+m'^2)^2(1+5(Z^2 + Z^4))\\
    & -4J^3(F_{2}^2+m'+m'^2)Z(1+5(Z^2+Z^4))+(-8+J^2(1-2F_{2}^2+2m'(1+m'))+4JZ)\cos(\varphi))\sin(k_{F}a)\\
    & -4Z^2(1+2Z^2)(-16Z+J^2(-4Z+(F_{2}^2+m'+m'^2)(8Z+J(4+3J(F_{2}^2+m'+m'^2)Z-4Z^2))))\sin(2k_{F}a)\\
    & -2Z^3(16-48Z^2+16JZ(-3+Z^2)+4J^3(F_{2}^2+m'+m'^2)Z(-3+Z^2)-J^4(F_{2}^2+m'+m'^2)^2(-1+3Z^2)\\
    & -4J^2(-1+2F_{2}^2+2m'(1+m'))(-1+3Z^2))\sin(3k_{F}a))\\\nonumber
\end{split}\\\nonumber
\begin{split}
B(\varphi)={}& -2J^2e^{6ik_{F}a}(-64F_{2}^4J^2(1+6(Z^2+Z^4))-3(1+2m')^2(32(Z^2+Z^4)+J^2(1+6(Z^2+Z^4)))+4F_{2}^2(-J^2(5\\
    & +4m'(1+m'))(1+6(Z^2+Z^4))-16(1+8(Z^2+Z^4)))+4J^2\cos(\varphi)+16F_{2}^2J^2\cos(\varphi)-64F_{2}^4J^2\cos(\varphi)\\
    & +16J^2m'\cos(\varphi)+16J^2m'^2\cos(\varphi)+128Z^2\cos(\varphi)+512F_{2}^2Z^2\cos(\varphi)+24J^2Z^2\cos(\varphi)+96F_{2}^2J^2Z^2\cos(\varphi)\\
    & -384F_{2}^4J^2Z^2\cos(\varphi)+512m'Z^2\cos(\varphi)+96J^2m'Z^2\cos(\varphi)+512m'^2Z^2\cos(\varphi)+96J^2m'^2Z^2\cos(\varphi)\\ 
    & +128Z^4\cos(\varphi)+512F_{2}^2Z^4\cos(\varphi)+24J^2Z^4\cos(\varphi)+96F_{2}^2J^2Z^4\cos(\varphi)-384F_{2}^4J^2Z^4\cos(\varphi)\\ 
    & +512m'Z^4\cos(\varphi)+96J^2m'Z^4\cos(\varphi)+512m'^2Z^4\cos(\varphi)+96J^2m'^2Z^4\cos(\varphi)-8JZ(1+2Z^2)\cos(k_{F}a)\\
    & (-8F_{2}^2-(1+2m')^2+(1+2m')^2\cos(\varphi))(-4+JZ+4F_{2}^2JZ+(4+(-1+4F_{2}^2)JZ)\cos(\varphi))+4Z^2\cos(2k_{F}a)\\
    & (-8F_{2}^2-(1+2m')^2+(1+2m')^2\cos(\varphi))(16-16JZ-16Z^2+(1+4F_{2}^2)J^2(-1+Z^2)+(16JZ+16(-1+Z^2)\\
    & +(-1+4F_{2}^2)J^2(-1+Z^2))\cos(\varphi))+64F_{2}^2\cos(2\varphi)J^2\cos(2\varphi)+4F_{2}^2J^2\cos(2\varphi)-4J^2m'\cos(2\varphi)\\
    & +16F_{2}^2J^2m'\cos(2\varphi)-4J^2m'^2\cos(2\varphi)+16F_{2}^2J^2m'^2\cos(2\varphi)-32Z^2\cos(2\varphi)-6J^2Z^2\cos(2\varphi)\\
    & +24F_{2}^2J^2Z^2\cos(2\varphi)-128m'Z^2\cos(2\varphi)-24J^2m'Z^2\cos(2\varphi)+96F_{2}^2J^2m'Z^2\cos(2\varphi)-128m'^2Z^2\cos(2\varphi)\\
    & -24J^2m'^2Z^2\cos(2\varphi)+96F_{2}^2J^2m'^2Z^2\cos(2\varphi)-32Z^4\cos(2\varphi)-6J^2Z^4\cos(2\varphi)+24F_{2}^2J^2Z^4\cos(2\varphi)\\
    & -128m'Z^4\cos(2\varphi)-24J^2m'Z^4\cos(2\varphi)+96F_{2}^2J^2m'Z^4\cos(2\varphi)-128m'^2Z^4\cos(2\varphi)-24J^2m'^2Z^4\cos(2\varphi)\\ 
    & +96F_{2}^2J^2m'^2Z^4\cos(2\varphi)+8JZ(1+2Z^2)(-8F_{2}^2-(1+2m')^2+(1+2m')^2\cos(\varphi))(J+4F_{2}^2J+4Z+(-J\\
    & +4F_{2}^2J-4Z)\cos(\varphi))\sin(k_{F}a)-8Z^2(-8F_{2}^2-(1+2m')^2+(1+2m')^2\cos(\varphi))(-16Z+J(-4+Z(J+4F_{2}^2J+4Z))\\ 
    & +(16Z+J(4+(-1+4F_{2}^2)JZ-4Z^2))\cos(\varphi))\sin(2k_{F}a))\\\nonumber 
\end{split}\\\nonumber
\begin{split}
C={}& e^{3ik_{F}a}(-1-2Z^2+2Z^2\cos(k_{F}a)-2Z\sin(k_{F}a))(4J^2(1+2F_{2}^2+2m'(1+m')+2Z^2-4(F_{2}^2+m'+m'^2)Z^2-2(-1+2F_{2}^2\\
    & +2m'(1+m'))Z^4)+16(1+6(Z^2+Z^4))+J^4(F_{2}^2+m'+m'^2)^2(1+6(Z^2+Z^4))+2Z(-2(-4+J^2(F_{2}^2+m'+m'^2))\\ 
    & (1+2Z^2)(4Z+J(2+J(F_{2}^2+m'+m'^2)Z))\cos(k_{F}a)+Z(4(-1+Z)+J(J(F_{2}^2+m'+m'^2)(-1 + Z)+2(1 + Z)))\\
    & (4(1+Z)+J(2-2Z+J(F_{2}^2+m'+m'^2)(1+Z)))\cos(2k_{F}a)+2(-4+J^2(F_{2}^2+m'+m'^2))(4+J(J(F_{2}^2+m'+m'^2)\\
    & -2Z))(1+2Z^2)\sin(k_{F}a)+2Z(-4+J(-J(F_{2}^2+m'+m'^2)+2Z))(4Z+J(2+J(F_{2}^2+m'+m'^2)Z))\sin(2k_{F}a)))\\\nonumber 
\end{split}\nonumber
\end{align}
\section{Table}
To study the effect of high spin states of HSM on the Josephson supercurrent (Eq.~(8)) in section 11 of our paper, we provide spin flip probability ($F_{2}$) values of the HSM for different $S$ and $m'$ in a tabular format.
\begin{table}[h]
\tiny
\caption{Spin flip probability ($F_{2}$) values of the HSM for different $S$ and $m'$}
\begin{tabular}[t]{ c  c  c }
\toprule%
$S$\hspace{10ex} & $m'$\hspace{10ex} & $F_{2}$\\\hline \vspace{5px}
$\frac{1}{2}$\hspace{10ex} &$-\frac{1}{2}$\hspace{10ex} &$1$\\\vspace{5px}
                           &$\frac{1}{2}$\hspace{9ex}   &$0$\\\hline\vspace{5px}
$\frac{3}{2}$\hspace{10ex} &$-\frac{3}{2}$\hspace{10ex} &$\sqrt{3}$\\\vspace{5px}
                           &$-\frac{1}{2}$\hspace{10ex} &$2$\\\vspace{5px}
                           &$\frac{1}{2}$\hspace{9ex}   &$\sqrt{3}$\\\vspace{5px}
                           &$\frac{3}{2}$\hspace{9ex}   &$0$\\\hline\vspace{5px}
$\frac{5}{2}$\hspace{10ex} &$-\frac{5}{2}$\hspace{10ex} &$\sqrt{5}$\\\vspace{5px}
              &$-\frac{3}{2}$\hspace{10ex} &$2\sqrt{2}$\\\vspace{5px}
              &$-\frac{1}{2}$\hspace{10ex} &$3$\\\vspace{5px}
              &$\frac{1}{2}$\hspace{9ex}   &$2\sqrt{2}$\\\vspace{5px}
              &$\frac{3}{2}$\hspace{9ex}   &$\sqrt{5}$\\\vspace{5px}
              &$\frac{5}{2}$\hspace{9ex}   &$0$\hspace{10ex}\\\hline \vspace{5px}
$\frac{7}{2}$\hspace{10ex} &$-\frac{7}{2}$\hspace{10ex} &$\sqrt{7}$\\\vspace{5px}
              &$-\frac{5}{2}$\hspace{10ex} &$2\sqrt{3}$\hspace{10ex}\\\vspace{5px}
              &$-\frac{3}{2}$\hspace{10ex} &$\sqrt{15}$\hspace{10ex}\\\vspace{5px}
              &$-\frac{1}{2}$\hspace{10ex} &$4$\hspace{10ex}\\\vspace{5px}
              &$\frac{1}{2}$\hspace{9ex}   &$\sqrt{15}$\hspace{10ex}\\\vspace{5px}
              &$\frac{3}{2}$\hspace{9ex}   &$2\sqrt{3}$\hspace{10ex}\\\vspace{5px}
              &$\frac{5}{2}$\hspace{9ex}   &$\sqrt{7}$\hspace{10ex}\\\vspace{5px}
              &$\frac{7}{2}$\hspace{9ex}   &$0$\hspace{10ex}\\\hline \vspace{5px}
$\frac{9}{2}$\hspace{10ex} &$-\frac{9}{2}$\hspace{10ex} &$3$\\\vspace{5px}
                           &$-\frac{7}{2}$\hspace{10ex} &$4$\\\vspace{5px}
                           &$-\frac{5}{2}$\hspace{10ex} &$\sqrt{21}$\\\vspace{5px}
                           &$-\frac{3}{2}$\hspace{10ex} &$2\sqrt{6}$\\\vspace{5px}
                           &$-\frac{1}{2}$\hspace{10ex} &$5$\\\vspace{5px}
                           &$\frac{1}{2}$\hspace{9ex}   &$2\sqrt{6}$\\\vspace{5px}
                           &$\frac{3}{2}$\hspace{9ex}   &$\sqrt{21}$\\\vspace{5px}
                           &$\frac{5}{2}$\hspace{9ex}   &$4$\\\vspace{5px}
                           &$\frac{7}{2}$\hspace{9ex}   &$3$\\\vspace{5px}
                           &$\frac{9}{2}$\hspace{9ex}   &$0$\\\hline
\end{tabular}
\hfill
\begin{tabular}[t]{ c  c  c }
\toprule%
$S$\hspace{10ex} & $m'$\hspace{10ex} & $F_{2}$\\\hline \vspace{5px}
$\frac{11}{2}$\hspace{10ex} &$-\frac{11}{2}$\hspace{10ex} &$\sqrt{11}$\\\vspace{5px}
                            &$-\frac{9}{2}$\hspace{10ex}  &$2\sqrt{5}$\\\vspace{5px}
                            &$-\frac{7}{2}$\hspace{10ex}  &$3\sqrt{3}$\\\vspace{5px}
                            &$-\frac{5}{2}$\hspace{10ex}  &$4\sqrt{2}$\\\vspace{5px}
                            &$-\frac{3}{2}$\hspace{10ex}  &$\sqrt{35}$\\\vspace{5px}
                            &$-\frac{1}{2}$\hspace{10ex}  &$6$\\\vspace{5px}
                            &$\frac{1}{2}$\hspace{9ex}    &$\sqrt{35}$\\\vspace{5px}
                            &$\frac{3}{2}$\hspace{9ex}    &$4\sqrt{2}$\\\vspace{5px}
                            &$\frac{5}{2}$\hspace{9ex}    &$3\sqrt{3}$\\\vspace{5px}
                            &$\frac{7}{2}$\hspace{9ex}    &$2\sqrt{5}$\\\vspace{5px}
                            &$\frac{9}{2}$\hspace{9ex}    &$\sqrt{11}$\\\vspace{5px}
                            &$\frac{11}{2}$\hspace{9ex}   &$0$\hspace{10ex}\\\hline \vspace{5px}
$\frac{13}{2}$\hspace{10ex} &$-\frac{13}{2}$\hspace{10ex} &$\sqrt{13}$\\\vspace{5px}
                            &$-\frac{11}{2}$\hspace{10ex} &$2\sqrt{6}$\hspace{10ex}\\\vspace{5px}
                            &$-\frac{9}{2}$\hspace{10ex}  &$\sqrt{33}$\hspace{10ex}\\\vspace{5px}
                            &$-\frac{7}{2}$\hspace{10ex}  &$2\sqrt{10}$\hspace{10ex}\\\vspace{5px}
                            &$-\frac{5}{2}$\hspace{10ex}  &$3\sqrt{5}$\hspace{10ex}\\\vspace{5px}
                            &$-\frac{3}{2}$\hspace{10ex}  &$4\sqrt{3}$\hspace{10ex}\\\vspace{5px}
                            &$-\frac{1}{2}$\hspace{10ex}  &$7$\hspace{10ex}\\\vspace{5px}
                            &$\frac{1}{2}$\hspace{9ex}    &$4\sqrt{3}$\hspace{10ex}\\\vspace{5px}
                            &$\frac{3}{2}$\hspace{9ex}    &$3\sqrt{5}$\\\vspace{5px}
                            &$\frac{5}{2}$\hspace{9ex}    &$2\sqrt{10}$\\\vspace{5px}
                            &$\frac{7}{2}$\hspace{9ex}    &$\sqrt{33}$\\\vspace{5px}
                            &$\frac{9}{2}$\hspace{9ex}    &$2\sqrt{6}$\\\vspace{5px}
                            &$\frac{11}{2}$\hspace{9ex}   &$\sqrt{13}$\\\vspace{5px}
                            &$\frac{13}{2}$\hspace{9ex}   &$0$\\\hline\vspace{5px}
\end{tabular}
\hfill
\begin{tabular}[t]{ c  c  c }
\toprule%
$S$\hspace{10ex} & $m'$\hspace{10ex} & $F_{2}$\\\hline \vspace{5px}
$\frac{15}{2}$\hspace{10ex} &$-\frac{15}{2}$\hspace{10ex} &$\sqrt{15}$\\\vspace{5px}
                            &$-\frac{13}{2}$\hspace{10ex} &$2\sqrt{7}$\\\vspace{5px}
                            &$-\frac{11}{2}$\hspace{10ex} &$\sqrt{39}$\\\vspace{5px}
                            &$-\frac{9}{2}$\hspace{10ex}  &$4\sqrt{3}$\\\vspace{5px}
                            &$-\frac{7}{2}$\hspace{10ex}  &$\sqrt{55}$\\\vspace{5px}
                            &$-\frac{5}{2}$\hspace{10ex}  &$2\sqrt{15}$\\\vspace{5px}
                            &$-\frac{3}{2}$\hspace{10ex}  &$3\sqrt{7}$\\\vspace{5px}
                            &$-\frac{1}{2}$\hspace{10ex}  &$8$\\\vspace{5px}
                            &$\frac{1}{2}$\hspace{9ex}    &$3\sqrt{7}$\\\vspace{5px}
                            &$\frac{3}{2}$\hspace{9ex}    &$2\sqrt{15}$\\\vspace{5px}
                            &$\frac{5}{2}$\hspace{9ex}    &$\sqrt{55}$\\\vspace{5px}
                            &$\frac{7}{2}$\hspace{9ex}    &$4\sqrt{3}$\\\vspace{5px}
                            &$\frac{9}{2}$\hspace{9ex}    &$\sqrt{39}$\\\vspace{5px}
                            &$\frac{11}{2}$\hspace{9ex}   &$2\sqrt{7}$\\\vspace{5px}
                            &$\frac{13}{2}$\hspace{9ex}   &$\sqrt{15}$\\\vspace{5px}
                            &$\frac{15}{2}$\hspace{9ex}   &$0$\hspace{10ex}\\\hline\vspace{5px}
$\frac{17}{2}$\hspace{10ex} &$-\frac{17}{2}$\hspace{10ex} &$\sqrt{17}$\\\vspace{5px}
                            &$-\frac{15}{2}$\hspace{10ex} &$4\sqrt{2}$\\\vspace{5px}
                            &$-\frac{13}{2}$\hspace{10ex} &$3\sqrt{5}$\\\vspace{5px}
                            &$-\frac{11}{2}$\hspace{10ex} &$2\sqrt{14}$\\\vspace{5px}
                            &$-\frac{9}{2}$\hspace{10ex}  &$\sqrt{65}$\\\vspace{5px}
                            &$-\frac{7}{2}$\hspace{10ex}  &$6\sqrt{2}$\\\vspace{5px}
                            &$-\frac{5}{2}$\hspace{10ex}  &$\sqrt{77}$\\\vspace{5px}
                            &$-\frac{3}{2}$\hspace{10ex}  &$4\sqrt{5}$\\\vspace{5px}
                            &$-\frac{1}{2}$\hspace{10ex}  &$9$\\\vspace{5px}
                            &$\frac{1}{2}$\hspace{9ex}    &$4\sqrt{5}$\\\vspace{5px}
                            &$\frac{3}{2}$\hspace{9ex}    &$\sqrt{77}$\\\vspace{5px}
                            &$\frac{5}{2}$\hspace{9ex}    &$6\sqrt{2}$\\\vspace{5px}
                            &$\frac{7}{2}$\hspace{9ex}    &$\sqrt{65}$\\\vspace{5px}
                            &$\frac{9}{2}$\hspace{9ex}    &$2\sqrt{14}$\\\vspace{5px}
                            &$\frac{11}{2}$\hspace{9ex}   &$3\sqrt{5}$\\\vspace{5px}
                            &$\frac{13}{2}$\hspace{9ex}   &$4\sqrt{2}$\\\vspace{5px}
                            &$\frac{15}{2}$\hspace{9ex}   &$\sqrt{17}$\\\vspace{5px}
                            &$\frac{17}{2}$\hspace{9ex}   &$0$\\\hline
\end{tabular}
\hfill
\begin{tabular}[t]{ c  c  c }
\toprule%
$S$\hspace{10ex} & $m'$\hspace{10ex} & $F_{2}$\\\hline \vspace{5px}
$\frac{19}{2}$\hspace{10ex} &$-\frac{19}{2}$\hspace{10ex}  &$\sqrt{19}$\\\vspace{5px}
                            &$-\frac{17}{2}$\hspace{10ex}  &$6$\\\vspace{5px}
                            &$-\frac{15}{2}$\hspace{10ex}  &$\sqrt{51}$\\\vspace{5px}
                            &$-\frac{13}{2}$\hspace{10ex}  &$8$\\\vspace{5px}
                            &$-\frac{11}{2}$\hspace{10ex}  &$5\sqrt{3}$\\\vspace{5px}
                            &$-\frac{9}{2}$\hspace{10ex}   &$2\sqrt{21}$\\\vspace{5px}
                            &$-\frac{7}{2}$\hspace{10ex}   &$\sqrt{91}$\\\vspace{5px}
                            &$-\frac{5}{2}$\hspace{10ex}   &$4\sqrt{6}$\\\vspace{5px}
                            &$-\frac{3}{2}$\hspace{10ex}   &$3\sqrt{11}$\\\vspace{5px}
                            &$-\frac{1}{2}$\hspace{10ex}   &$10$\\\vspace{5px}
                            &$\frac{1}{2}$\hspace{9ex}     &$3\sqrt{11}$\\\vspace{5px}
                            &$\frac{3}{2}$\hspace{9ex}     &$4\sqrt{6}$\\\vspace{5px}
                            &$\frac{5}{2}$\hspace{9ex}     &$\sqrt{91}$\\\vspace{5px}
                            &$\frac{7}{2}$\hspace{9ex}     &$2\sqrt{21}$\\\vspace{5px}
                            &$\frac{9}{2}$\hspace{9ex}     &$5\sqrt{3}$\\\vspace{5px}
                            &$\frac{11}{2}$\hspace{9ex}    &$8$\\\vspace{5px}
                            &$\frac{13}{2}$\hspace{9ex}    &$\sqrt{51}$\\\vspace{5px}
                            &$\frac{15}{2}$\hspace{9ex}    &$6$\\\vspace{5px}
                            &$\frac{17}{2}$\hspace{9ex}    &$\sqrt{19}$\\\vspace{5px}
                            &$\frac{19}{2}$\hspace{9ex}    &$0$\\\hline
\end{tabular}
\end{table}

\end{document}